\PassOptionsToPackage{table}{xcolor}
\documentclass[sigconf, balance=false]{acmart}
\usepackage{popets}

\setcopyright{popets}
\copyrightyear{YYYY}

\acmYear{YYYY}
\acmVolume{YYYY}
\acmNumber{X}
\acmDOI{XXXXXXX.XXXXXXX}
\acmISBN{}
\acmConference{Proceedings on Privacy Enhancing Technologies}
\usepackage{tikz}
\usepackage{amsmath}
\usepackage{enumitem}
\usepackage{enumerate}
\usepackage{wasysym}
\usepackage{subcaption}
\usepackage{xcolor}
\usepackage{booktabs}
\usepackage{makecell}
\usepackage{hyperref}
\usepackage{adjustbox}
\usepackage{todonotes}
\usepackage{tcolorbox}
\usepackage{tikz}
\usepackage[framemethod=TikZ]{mdframed}
\usepackage{wrapfig}

\newcommand{\quoteFrame}[1]{%
    \begin{mdframed}[leftline=true, topline=false, bottomline=false, rightline=false, linewidth=2pt, linecolor=gray, innertopmargin=1pt, innerbottommargin=1pt] 
       #1
    \end{mdframed}
}

\title[Privacy Bills of Materials (\texttt{PriBOM})]{Privacy Bills of Materials (\texttt{PriBOM}): A Transparent Privacy Information Inventory for Collaborative Privacy Notice Generation in Mobile App Development}

\author{Zhen Tao}
\affiliation{%
  \institution{Australian National University \& CSIRO's Data61}
  \country{Australia}
}

\author{Shidong Pan}
\authornote{Co-first author with equal contribution. shidong.pan@nyu.edu, Computer Science Department at New York University \& Columbia Law School.}
\affiliation{%
  \institution{New York University \& Columbia University}
  \country{USA}
}

\author{Zhenchang Xing}
\affiliation{%
  \institution{CSIRO's Data61}
  \country{Australia}
}

\author{Xiaoyu Sun}
\affiliation{%
  \institution{Australian National University}
  \country{Australia}
}

\author{Omar Haggag}
\affiliation{%
  \institution{Monash University}
  \country{Australia}
}

\author{John Grundy}
\affiliation{%
  \institution{Monash University}
  \country{Australia}
}

\author{Jingjie Li}
\authornote{Co-senior author.}
\affiliation{%
  \institution{University of Edinburgh}
  \country{UK}
}

\author{Liming Zhu}
\affiliation{%
  \institution{CSIRO's Data61 \& School of CSE, UNSW}
  \country{Australia}
}

\begin{document}

\renewcommand{\shortauthors}{Tao et al.}

\begin{abstract}
Privacy regulations mandate that developers must provide authentic and comprehensive privacy notices, e.g., privacy policies or labels, to inform users of their apps’ privacy practices. 
However, due to a lack of knowledge of privacy requirements, developers often struggle to create accurate privacy notices, especially for sophisticated mobile apps with complex features and in crowded development teams. 
To address these challenges, we introduce \texttt{PriBOM} (Privacy Bills of Materials), a systematic software engineering approach that leverages different development team roles to better capture and coordinate mobile app privacy information. 
\texttt{PriBOM} facilitates transparency-centric privacy documentation and specific privacy notice creation,  enabling traceability and trackability of privacy practices. We present a pre-fill of \texttt{PriBOM} based on static analysis and privacy notice analysis techniques.
We explore the perceived usefulness of \texttt{PriBOM} through a human evaluation with 150 diverse participants. 
The role of \texttt{PriBOM} in enhancing privacy-related communication is well received with 83.33\% agreement, suggesting that \texttt{PriBOM} could serve as a significant solution for providing privacy support in DevOps for mobile apps.
\end{abstract}

\maketitle


\section{Introduction}
\label{sec_intro}
Due to functional, analytical, and advertising needs, mobile application developers are increasingly expanding their collection and use of users' personal information and other privacy-related data. 
Many privacy regulations, such as the General Data Protection Regulation (GDPR)~\cite{GDPR}, the California Consumer Privacy Act (CCPA)~\cite{CCPA}, and the Australian Privacy Principles (APP)~\cite{APPs}, require developers to provide \emph{authentic} and \emph{understandable} privacy notices to inform users of the app's privacy practices. Privacy policy is the most prevailing format of privacy notices to mobile application users~\cite{adams2020agreeing, bui2023detection, flavian2006consumer, caramujo2015analyzing, perez2018review, harkous2018polisis, kemp2020concealed}. In pursuit of higher readability and conciseness, mainstream application stores in the market have also recently required app developers to remind users of app's potential privacy data practices in the form of privacy nutrition labels (\textit{a.k.a.} privacy label)~\cite{kelley2009nutrition, kelley2010standardizing, kelley2013privacy}, e.g. the Data Safety Sections (DSS) in Google Play and the Apple Privacy Labels (APL) in Apple's App Store.

However, numerous studies~\cite{balebako2014improving, balebako2014privacy, li2022understanding, pan2023toward} have shown that existing app privacy notices are often problematic, as they fail to authentically align with the actual data practices of apps. 
While non-comprehensive and inaccurate privacy notices could harm users' trustworthiness and violate privacy regulations, software developers face various challenges when providing authoritative privacy notices. 
Crafting a good privacy notice is complex, requiring not only legal knowledge but also a fundamental understanding of the app's various functions and features. 
Developers commonly lack training or knowledge in privacy and legal fields~\cite{li2022understanding, li2018coconut, li2021developers}, and often hold a passive or even negative attitude towards privacy factors during development~\cite{li2022understanding, li2021developers, lee2024don}. 
Additionally, even though legal teams are mainly responsible for the privacy notice documentation, they are naturally not acquainted with mobile app technical details.
Such numerous challenges have promoted the development of assistance tools in generating privacy notices.


There are three types of mainstream tools to help the development team generate privacy notices: Online Automated Privacy Policy Generators (APPGs)~\cite{pantrap, Iubenda, Appprivacypolicygenerator, Termly, Privacypolicies, Privacypolicyonline}, Code-based Privacy Policy Generators (CPPGs)~\cite{yu2018ppchecker, yu2015autoppg, zimmeck2021privacyflash} and the recently emerged Code-based IDE Plugins
(CIDEPs)~\cite{li2018coconut, li2021honeysuckle, li2024matcha}.
Although these tools are useful and widely adopted, most are not applicable to sophisticated mobile apps with complex features and in crowded development teams. 
A small change in a basic function could require significant effort to accurately reflect the privacy practices in privacy notices.
Concurrently, thousands of such modifications will occur in the software development and maintenance process, making it impossible to manually track all those privacy practice changes, ultimately leading to problematic privacy notices.
These issues are further magnified in the development under multi-role collaboration. The legal team often feel like ``lone wolves'' carrying the company's privacy program alone~\cite{LinkedIn}, given that other roles are minimally involved in managing privacy.
Thus, existing generators are far from enough, a systematic and collaborative software engineering solution involving various roles is pressingly needed.

\begin{figure}[t]
  \centering
  \includegraphics[width=.8\linewidth]{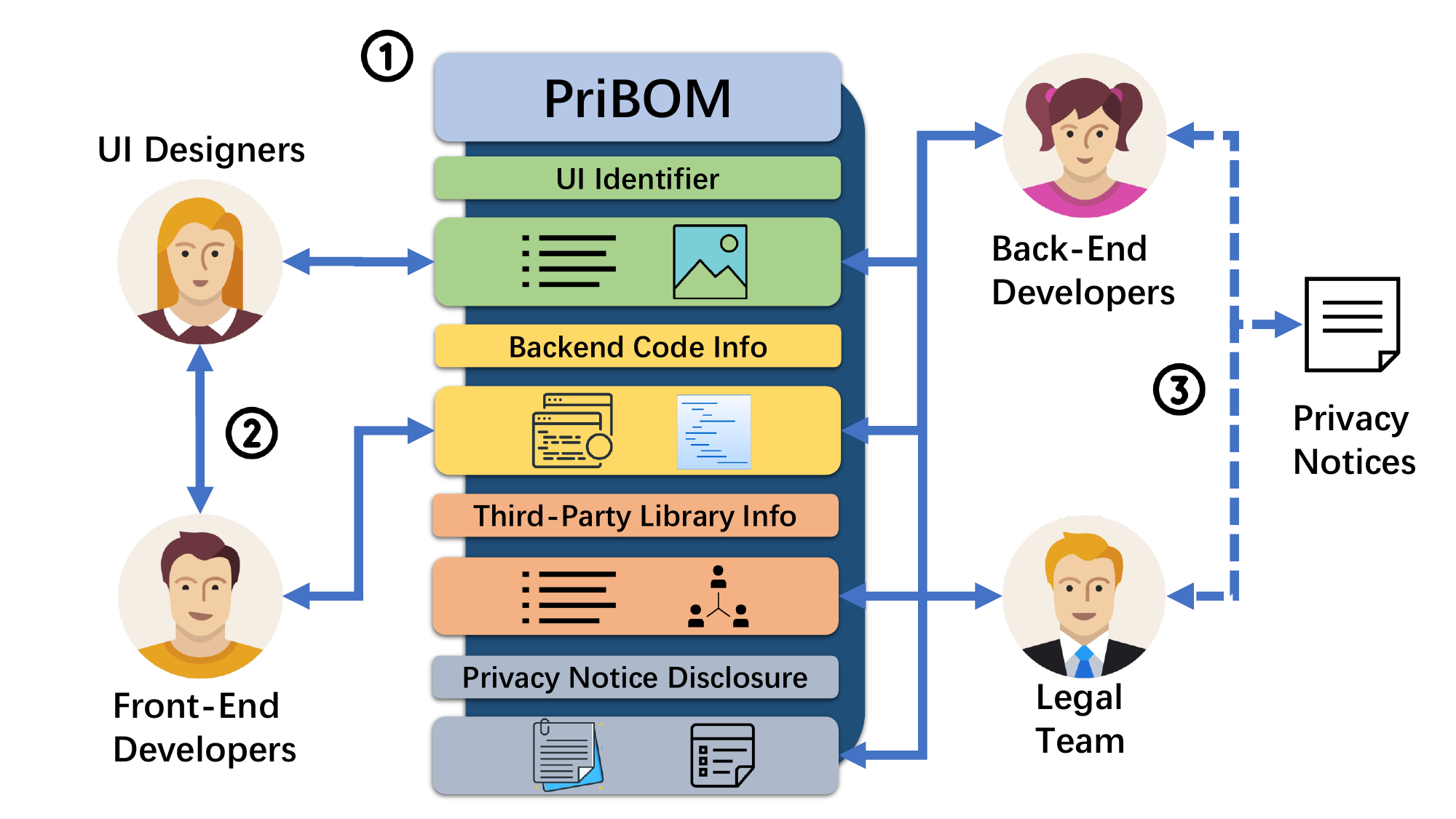}

  \caption{Use cases of \texttt{PriBOM}. (1) A privacy information inventory indexed by UI widgets, providing transparent privacy documentation. (2) A privacy communication platform between different roles in the development team. (3) A systematic solution for collaborative privacy notice generation.
  }
  \label{fig_use_case}
\end{figure}

To tackle this challenge, we propose \texttt{PriBOM}
(\underline{Pri}vacy \underline{B}ills \underline{o}f \underline{M}aterials), a systematic approach that stores privacy practices in a structured manner and facilitates the transparent, collaborative, and accurate generation of privacy notices.
The concept of \texttt{PriBOM} is inspired by the rising of Software Bills of Materials (SBOM).
Figure~\ref{fig_use_case} illustrates the use cases of \texttt{PriBOM}.
In this paper, we first retrospect the development history of privacy notice and generation tools and then conduct a literature review as a formative study. 
The study reveals three major challenges to software privacy factors encountered by developers in DevOps: 1) Privacy Knowledge Absence, 2) Limited Technical Knowledge, and 3) Unfriendly Organizational Environment.
We then scrutinize how previous privacy notice generation tools fail to mitigate those challenges and highlight the necessity for a revolutionary solution.
After, we introduce the motivation and the design of \texttt{PriBOM} in detail.

Focusing on facilitating collaborative privacy notice generation in mobile app development, \texttt{PriBOM} is a table-like privacy information inventory indexed by UI widget,
documenting information regarding 1) UI Identifier, 2) Codebase and Permission, 3) Third-Party Library and 4) Privacy Notice Disclosure.
Table~\ref{tab:pribom_format} presents the format of \texttt{PriBOM}.
The UI widget serves as both visual elements and key components in functionality and data handling, therefore it is the pivot to synergistically connect different roles of developers on privacy-related communication.
Additionally, we present a pre-fill of \texttt{PriBOM} based on cutting-edge static analysis and privacy notice analysis techniques, demonstrating its practicability.
Furthermore, we conducted a usability evaluation of \texttt{PriBOM} through a survey of 150 participants. 
By using a survey, we also aim to prompt discussions and opinions around potential adaption to specific applications beyond only validating its usefulness.
The statements about design intuitiveness, content comprehension, and information relevance of \texttt{PriBOM} receive positive feedback,
underlined by the 85.3\%, 72\%, and 78.76\% agreement, respectively.
We observed differences in the perspectives held by different roles, with experience within the same role further shaping these viewpoints. 
Non-technical roles frequently highlighted \texttt{PriBOM}'s efficiency in streamlining workflows.
Lastly, we discussed potential enhancements and ongoing refinements to ensure its practical adaptation and its alignment with real-world demands across roles.
The implementation and survey questionnaire are available at:~\url{https://anonymous.4open.science/r/PriBOM-CE36}.

In summary, our
\texttt{PriBOM} provides a transparent privacy information inventory, greatly facilitating collaborative privacy notice generation. The key contributions are: 
\begin{itemize} [leftmargin=*, noitemsep, topsep=3pt]
\item To the best of our knowledge, we are the first to systematically summarise the privacy notice generation tools.
\item We introduce the concept of \texttt{PriBOM} (Privacy Bills of Materials) and propose a pre-fill for mobile app development.
\item We conduct a human evaluation to comprehsnsively assess the usefulness of \texttt{PriBOM}.
\end{itemize}



\section{Privacy Notice Generation for Mobile Apps}
\label{sec_background}



\subsection{Status Quo}




Although privacy policies have become the primary privacy notice approach for mobile applications~\cite{adams2020agreeing, bui2023detection, flavian2006consumer, caramujo2015analyzing, perez2018review, harkous2018polisis, kemp2020concealed}, their presentation and readability have always been criticized~\cite{mcdonald2008cost, JarniPrivacypolicy}. 
To improve the usability, Kelly et al.~\cite{kelley2009nutrition, kelley2010standardizing, kelley2013privacy} introduced the privacy nutrition labels, or privacy labels, designed to facilitate consumers’ understanding of how their information is collected and utilized in a concise and structured manner.
Privacy labels have been widely adopted by practitioners and have become a trend for conveying apps' privacy practices to end users. 
As required by privacy regulations (e.g., GDPR), providing accurate privacy notices is equally important as providing them in an inviting way.
Developers are responsible for creating accurate privacy notices~\cite{li2024matcha}, though various challenges and concerns have been discovered and raised~\cite{li2022understanding, balebako2014improving, balebako2014privacy}, even for big companies such as Google\footnote{\textit{In re Facebook, Inc. Internet Tracking Litigation,}
956 F. 3d 589 - Court of Appeals, 9th Circuit, 2020} and Facebook\footnote{\textit{In re Google Assistant Privacy Litigation,}
457 F. Supp. 3d 797 - Dist. Court, ND California}.
Failing to do so may cause serious legal consequences.
To respond to those challenges, prior efforts have been made to assist developers create privacy notices. 
Based on their inherent nature, these tools can be categorized into various groups, including Online Automated Privacy Policy Generators (APPGs)~\cite{pantrap, Iubenda, Appprivacypolicygenerator, Termly, Privacypolicies, Privacypolicyonline}, Code-based Privacy Policy Generators (CPPGs)~\cite{yu2018ppchecker, yu2015autoppg, zimmeck2021privacyflash} and Code-based IDE Plugins (CIDEPs)~\cite{li2018coconut, li2021honeysuckle, li2024matcha}. We introduce the evolution process of these tools and summarize their key features below.

%
\begin{figure} [!t]
\begin{subfigure}{.49\linewidth}
  \centering
  \includegraphics[width=.99\linewidth]{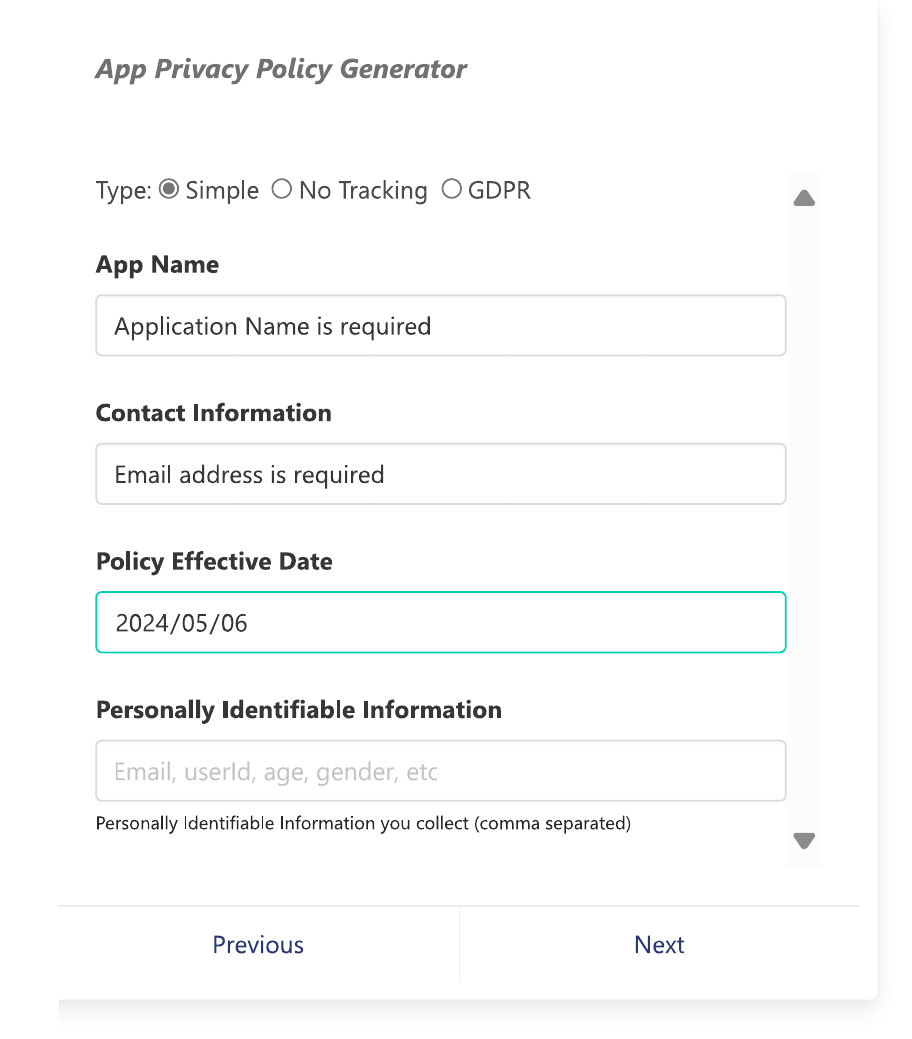}
  \caption[]{APPG PII}
  \label{fig_APPG_PII}
\end{subfigure}%
\begin{subfigure}{.49\linewidth}
  \centering
  \includegraphics[width=.99\linewidth]{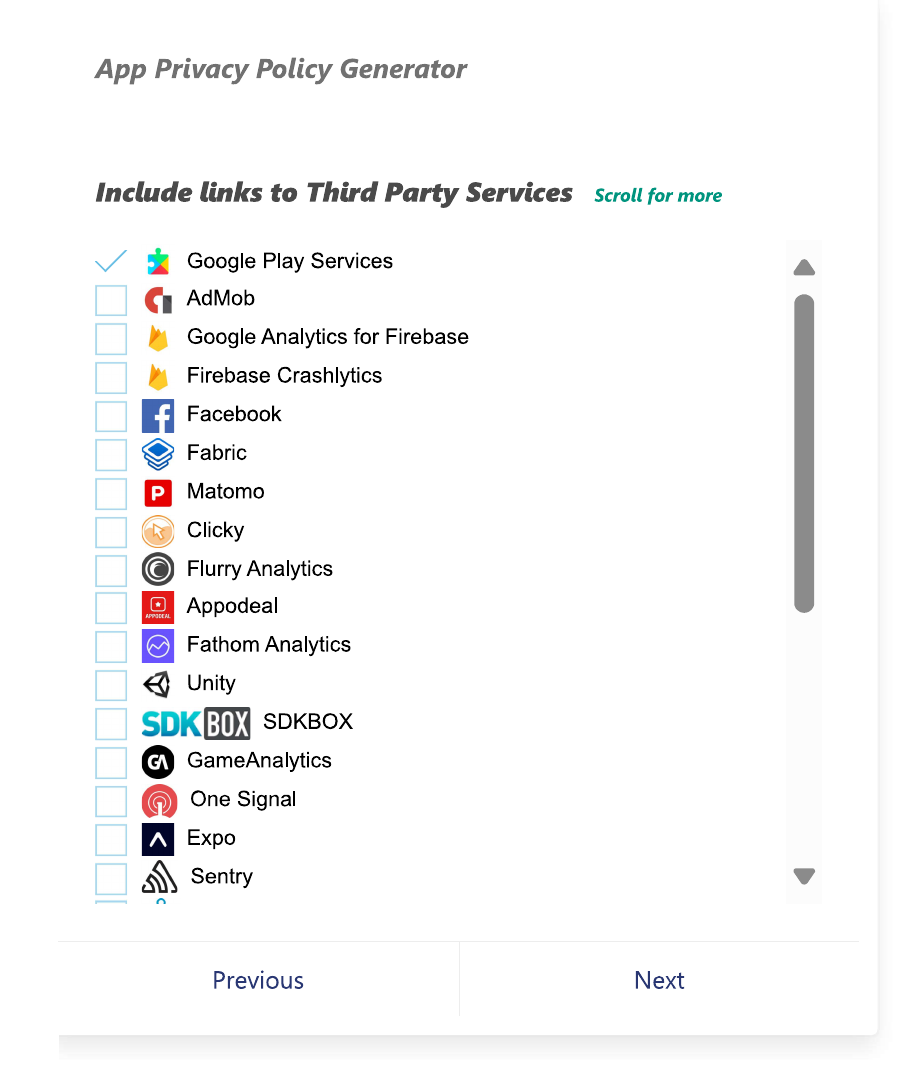}
  \caption{APPG TPL}
  \label{fig_APPG_TPL}
\end{subfigure}
\hfill
\begin{subfigure}{.98\linewidth}
  \centering
  \includegraphics[width=.99\linewidth]{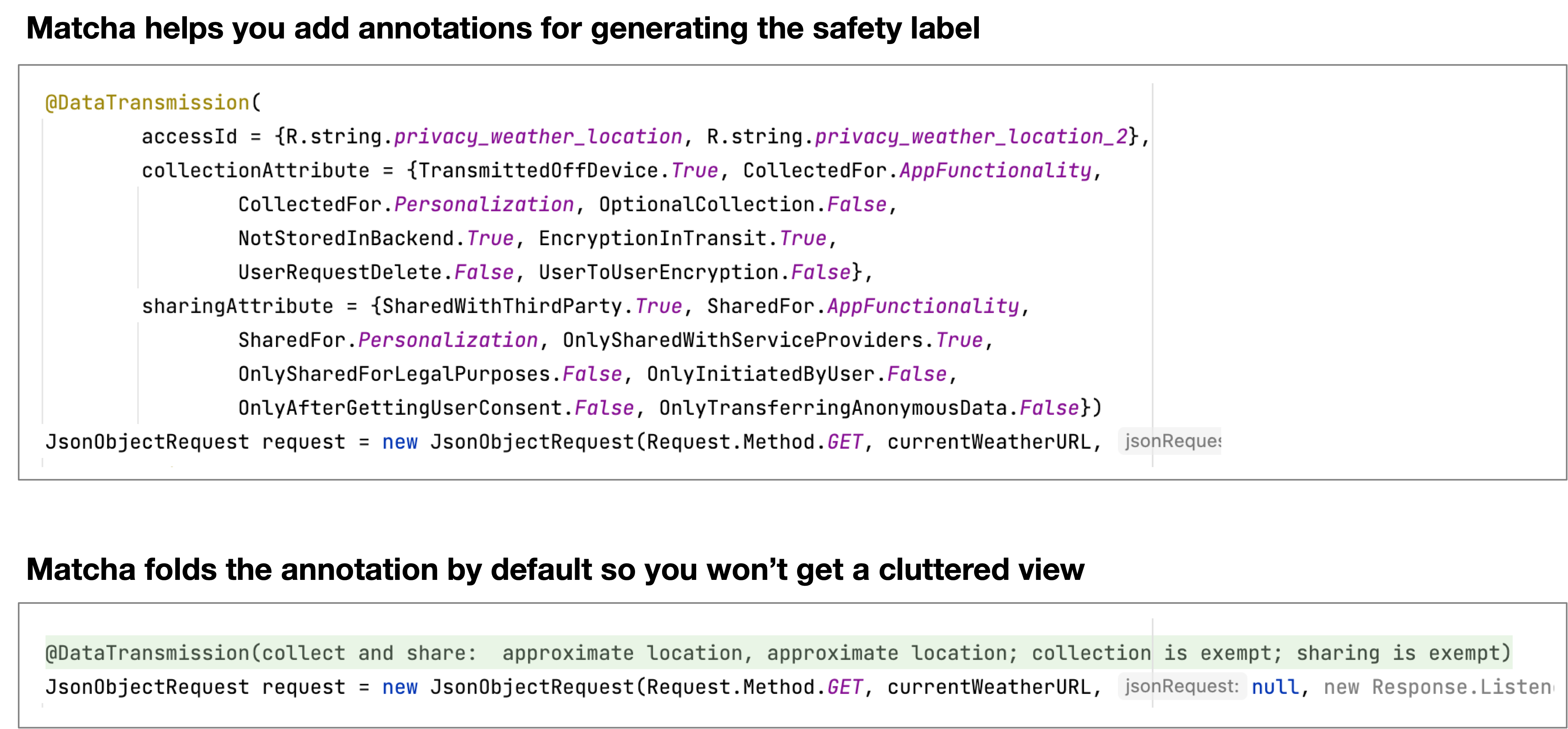}
  \caption{Matcha, a CIDEP proposed by Li et al.~\cite{li2024matcha}}
  \label{fig_CIDEP}
\end{subfigure}
\caption[Caption]{(a) and (b) are the interfaces of \cite{Appprivacypolicygenerator}, one of the most popular APPGs, according to Pan et al.~\cite{pantrap}, on collecting Personally Identifiable Information (PII) and Third-party Libraries (TPL) usages. (c) is a screenshot of Matcha~\cite{li2024matcha}, one of the Code-based IDE Plugins (CIDEP), from its JetBrains plugin page~\cite{MatchaJetBrainspluginpage}.}
\label{fig_tool}
\end{figure}
%

Most APPGs are questionnaire-based~\cite{pantrap} online tools that depend on developer-provided information to generate privacy policies, as shown in 
Figure~\ref{fig_APPG_PII} and Figure~\ref{fig_APPG_TPL}. 
Although APPGs can directly generate privacy policies, their qualities can be largely affected by developers' unartistic design flaws and inaccurate completion. 
As the privacy policy generation process by APPGs is completely disconnected from the original software development in DevOps, it is inevitable that developers cannot appropriately maneuver these tools.

CPPGs have been proposed to further make the privacy notice generation stage ``shift left'' in DevOps, i.e., in the earlier stages of the software development cycle.
Tools like AutoPPG~\cite{yu2015autoppg, yu2018ppchecker} and PrivacyFlash~\cite{zimmeck2021privacyflash} analyze privacy-related features contained in the source code of Android and iOS applications and generate privacy descriptions or notices based on code features. 
However, CPPGs face significant challenges, such as inherent complexity and low explainability, which hinder their adoption and effectiveness.
Also, CPPGs often fail to ensure compliance with high-level privacy regulations, particularly concerning non-functional requirements~\cite{pantrap}. 

Unlike generating complete privacy notices post-development, CIDEPs are integrated into the integrated development environment (IDE) to provide code privacy annotations for developers during the development process. Figure~\ref{fig_CIDEP} shows an example~\cite{li2024matcha} of such CIDEPs. These tools, such as Coconut~\cite{li2018coconut}, Honeysuckle~\cite{li2021honeysuckle} and Matcha~\cite{li2024matcha}, help developers add privacy annotations to provide information for privacy notice creation, thereby reducing common misunderstandings among developers and easing the creation of privacy notices.

\begin{table*}[!t]
\centering
  \caption{Summary of developer's privacy challenges and comparison of assistant tools along those challenges. "APPG" stands for Questionnaire-based Online Automated Privacy Policy Generators, "CPPG" stands for Code-based Privacy Policy Generators, "CIDEP" stands for Code-based IDE Plugins. \CIRCLE: addressed. \LEFTcircle: partially addressed. \Circle: not addressed.}
  \label{tab:dev_privacy_challenges}
  \resizebox{0.8\linewidth}{!}{%
  \begin{tabular}{l | c c c |c}
  \toprule
    \makecell[c]{\textbf{Privacy Challenges in Software Development Process}} & \textbf{APPG} & \textbf{CPPG} & \textbf{CIDEP} & \textbf{PriBOM}  \\
    \midrule

    \multicolumn{5}{l}{\textbf{[Challenge-1]} Privacy Knowledge Absence } \\ \midrule
    Misunderstanding of privacy terms~\cite{khandelwal2023unpacking, li2022understanding, li2018coconut, li2021developers, balebako2014privacy, tahaei2020understanding} & \LEFTcircle & \LEFTcircle & \LEFTcircle & \CIRCLE\\
    Lack of knowledge in privacy and legal field~\cite{khandelwal2023unpacking, li2022understanding, li2018coconut, li2021developers, balebako2014privacy, lee2024don, weir2020needs, kekulluouglu2023we} & \Circle & \Circle & \Circle & \LEFTcircle\\ 
    Update and iteration of privacy rules~\cite{khandelwal2023unpacking, li2021developers} & \Circle & \Circle & \CIRCLE & \CIRCLE\\ 

    \midrule
    \multicolumn{5}{l}{\textbf{[Challenge-2]} Limited Technical Knowledge} \\ \midrule
    Opacity of third-party libraries and resources~\cite{khandelwal2023unpacking, li2022understanding, li2018coconut, li2021developers, balebako2014privacy, seymour2023voice} & \Circle & \CIRCLE & \CIRCLE & \CIRCLE\\
    Complicated privacy notice creation process~\cite{khandelwal2023unpacking, li2022understanding} & \CIRCLE & \LEFTcircle & \Circle & \LEFTcircle\\
    Lack of awareness of privacy-preserving alternative implementations~\cite{li2018coconut, li2021developers, balebako2014privacy} & \Circle & \Circle & \LEFTcircle & \LEFTcircle\\
    Limited tool support in understanding data practices~\cite{li2018coconut, li2021developers, lee2024don, tahaei2021privacy, tahaei2020understanding} & \Circle & \LEFTcircle & \CIRCLE & \CIRCLE\\ 

    \midrule
    \multicolumn{5}{l}{\textbf{[Challenge-3]} Unfriendly Organizational Environment} \\ \midrule
    Not well-maintained privacy documentations~\cite{li2018coconut} & \Circle & \Circle & \LEFTcircle & \CIRCLE\\ 
    Negative and demotivated attitude towards privacy~\cite{li2022understanding, li2018coconut, li2021developers, balebako2014privacy, lee2024don, seymour2023voice, weir2020needs, kekulluouglu2023we, tahaei2021privacy} & \Circle & \Circle & \Circle & \LEFTcircle\\
    Lack of team, organization and platform support~\cite{li2022understanding, li2018coconut, lee2024don, seymour2023voice, weir2020needs, tahaei2020understanding} & \Circle & \Circle & \Circle & \CIRCLE\\ 
 
    \bottomrule
    
\end{tabular}
}%
\end{table*}
%
%

However, similar to APPGs and CPPGs, CIDEPs are also tailored for citizen developers or small teams in which only one or several developers are responsible for privacy notice generation.
Consequently, the usability and adaptability of these approaches are significantly constrained when applied to sophisticated mobile applications with complex features, typically developed by large teams. 
Therefore, we emphasize the need for a systematic solution integrated into DevOps that facilitates the collaborative generation of privacy notices. 
Understanding the specific challenges developers face is essential for developing an improved solution. Hence, we conducted a formative study to identify these challenges.

\subsection{Formative Study}
\label{sec_formative_study}


Developers often encounter privacy challenges as they attempt to build a thorough understanding of the privacy practices in the application and convert them into privacy notices~\cite{li2024matcha}. We first summarize the previous studies to analyze developers' challenges relevant to privacy aspects when they are developing applications or creating privacy notices. 



With the increasing attention paid to privacy issues in software development, many research works are dedicated to exposing the privacy challenges developers face in the increasingly complex software development process. To form a comprehensive understanding of the current status of privacy challenges faced by developers, we conduct a systematic literature review to summarize these privacy challenges and the research efforts on discovering and studying them. Our target venues are cybersecurity \textit{Big-Four}. We examined the titles of these papers from 2019 to 2024, and searched for keywords related to developer privacy challenges, such as \textit{``privacy'', ``developer'', ``development''}, and \textit{``privacy challenge''}, to conduct a preliminary selection of the papers. After obtaining the preliminarily selected papers, we manually checked the abstract and introduction of these papers to filter out papers that were not related to our topic. We then employed the snowballing method to comprehensively cover the relevant literature. First, we read through the related work section to identify related papers that are not published in target venues or do not include our keywords in the title. Second, we also examined papers that cited our selected papers. We eventually obtained 11 papers that discovered and studied privacy challenges faced by developers.



Table~\ref{tab:dev_privacy_challenges} presents a comprehensive overview of the multifaceted challenges developers encounter regarding privacy. To provide a structured approach to understanding these challenges, we categorize them into the following threefold:



\noindent \textbf{[Challenge-1] Privacy Knowledge Absence.} Such challenges pertain to the understanding of privacy within the development community. Developers often lack privacy knowledge~\cite{khandelwal2023unpacking, li2022understanding, li2018coconut, li2021developers, balebako2014privacy, lee2024don, weir2020needs, kekulluouglu2023we} due to the missing of formal privacy training, e.g. interpretation of privacy-related terminology, and may not be aware of alternative approaches that offer better privacy preservation~\cite{li2018coconut, li2021developers, balebako2014privacy}. Misunderstanding Privacy terms~\cite{khandelwal2023unpacking, li2022understanding, li2018coconut, li2021developers, balebako2014privacy, tahaei2020understanding} can lead to discrepancies between intended and implemented privacy practices, potentially causing privacy issues.

\noindent \textbf{[Challenge-2] Limited Technical Knowledge.} This kind of challenge involves privacy issues in the technical aspect, especially related to estimating a thorough understanding of the data practices. For example, developers often integrate Third-Party Libraries (TPLs) without a full understanding of their data handling practices~\cite{khandelwal2023unpacking, li2022understanding, li2018coconut, li2021developers, balebako2014privacy, seymour2023voice}, leading to a lack of transparency regarding functionality and privacy of external sources. Such opacity may cause developers to struggle to follow data minimization, which is a principle introduced by privacy regulations such as GDPR~\cite{GDPR} and formally defined by researchers~\cite{zhou2023policycomp}. Data minimization requires developers to only collect necessary personal data in relation to specific purposes. Although existing tools~\cite{li2018coconut, li2021developers, lee2024don, tahaei2021privacy, tahaei2020understanding} can provide support, such necessity checks still rely on manual checks by legal experts~\cite{zhou2023policycomp, wang2022privguard}.


\noindent \textbf{[Challenge-3] Unfriendly Organizational Environment.} These challenges reflect the broader environmental and motivational factors that influence privacy. The challenge of developers' negative and demotivated attitude towards privacy is mentioned by most studies~\cite{li2022understanding, li2018coconut, li2021developers, balebako2014privacy, lee2024don, seymour2023voice, weir2020needs, kekulluouglu2023we, tahaei2021privacy} and highlight the necessity of an urgent change of the attitude. There are mainly three reasons that contribute to this phenomenon. First, as privacy is considered to be a non-functional factor, trade-offs exist between privacy and other objectives like functionalities, user experience, business goals and model performance~\cite{lee2024don}. Second, the ownership of Privacy is uncertain. Developers may regard privacy as the responsibility of other personnel and not consider it in their own workflow~\cite{lee2024don}. Last, developers may find them in a workplace culture that may not prioritize privacy~\cite{tahaei2021privacy}. The absence of a supportive infrastructure for privacy within organizations can demotivate developers from pursuing stringent privacy standards. According to ~\cite{weir2020needs}, less than a quarter of developers have access to security experts. A disconnect between individual developers and decision-makers in the organization may also lead to negativeness towards privacy~\cite{lee2024don}.

\textbf{Performance of Existing Generation Tools.}
We compare the existing generation tools along the dimensions of summarized privacy challenges faced by developers. As presented in Table~\ref{tab:dev_privacy_challenges}, none of the existing tools can address all of these privacy challenges. First, these tools often assume citizen developers as their target audience, resulting in poor contributions to the privacy environment in large companies and organizations. Second, although CPPGs and CIDEPs can provide technical support for developers, they cannot improve the lack of privacy knowledge among developers. 

Given the inherent limitations of existing privacy notice generation tools, we argue that only a systematic software engineering solution can fundamentally tackle those problems.
Drawing from empirical observations and the formative study, we propose our design of \texttt{PriBOM} in collaborative mobile app development scenario.
\texttt{PriBOM} enhances transparency in privacy management and provides a platform for multi-role collaboration on privacy notice generation. We specifically introduce it in the next section.

\section{The \texttt{PriBOM} Approach}
\label{sec_PriBOM}


\begin{figure}[t!]
  \centering
  \includegraphics[width=.98\linewidth]{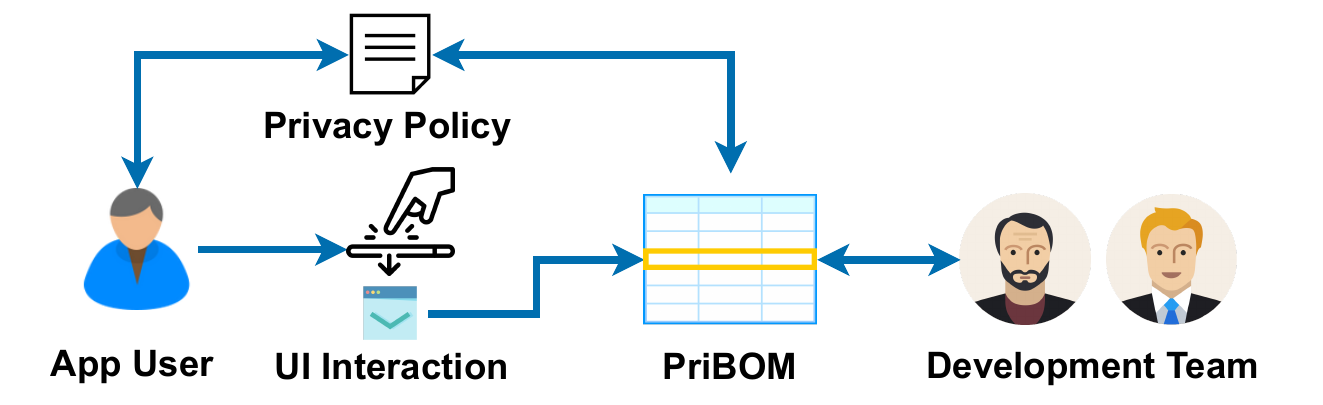}

  \vspace{-5pt}
  \caption{An overview of \texttt{PriBOM} in the usage scenario. The interaction between app users and UI components triggers data practices. These practices are disclosed to users through privacy notices such as the privacy policy. \texttt{PriBOM} helps development team create accurate privacy notices by documenting privacy information related to specific UI components.}
  \label{fig_PriBOM_flow}
\vspace{-5pt}
\end{figure}




\subsection{Motivation of \texttt{PriBOM}}
\label{sec_motv}


\noindent \textbf{Bill of Materials (BOM).}
To cater to increasingly complex and diverse software systems, Bill of Materials (BOM) has been introduced as the information inventory with a focus on the transparency of software materials. 
Proposed in the 2010s~\cite{SPDX}, SBOM has become a key strategy in response to the emerging risks in software vulnerabilities management, license compliance, and supply chain transparency~\cite{xia2023trust, BenefitsforSBOM}. 
SBOMs record the material details of the components in developing software, providing transparency, traceability and verifiability to software building and managing process~\cite{stalnaker2024boms, zahan2023software, mirakhorli2024landscape}. 
The US Presidential Executive Order (EO) 14028 on Improving
the Nation’s Cybersecurity\footnote{\href{https://www.nist.gov/itl/executive-order-14028-improving-nations-cybersecurity}{US Presidential Executive Order 14028}}
requires companies providing services to the US government to provide SBOMs. 
Approximately 78\% of the organizations worldwide would use SBOMs by 2022 and 88\% by 2023~\cite{stalnaker2024boms, StephenSoftware}. 
In addition to SBOM, the concept of BOM has been implemented into other forms, including HBOMs~\cite{HBOM} and FBOMs~\cite{AmyFBOM, AndreiFBOM, EliotFBOM} for hardware and firmware, DataBOMs~\cite{barclay2019towards} for datasets, and AIBOMs~\cite{BrianAIBOM,xia2023empirical} for AI models.
Consequently, we adapt a similar design for privacy information and introduce the concept of \texttt{PriBOM}.
\texttt{PriBOM} records software privacy information in a structured way and with a focus on UI, pulling different development roles to the same page. 
Specifically, \texttt{PriBOM} is indexed by the UI widget, including privacy information such as associated privacy permissions, data types, third-party libraries, and corresponding privacy notice disclosures. We will elaborate the design in Section~\ref{sec_bom_design}.



\noindent \textbf{UI Widgets - The Pivot of Mobile Privacy Communication.}
The UI widgets serve as both visual elements and key components in functionality and data handling, therefore it is the pivot to synergistically connect different roles on privacy communication.
First, UI components are fundamental in the software development process, which are not only visual elements but are often directly involved in the app's functionality and data handling. The transparency regarding what data is collected through specific UI components is one of the information that users want to know~\cite{pan2023seeprivacy}, especially when malicious design may lead to unwanted data collection and cause harm to users~\cite{brignull2010types, chen2023unveiling}. Moreover, UI represents the tangible interface where users interact with the application, making it a critical point that is associated with potential privacy issues. When conducting privacy-related tests, interaction with the UI widgets is also a crucial part of executing privacy checks~\cite{wang2018guileak, nan2015uipicker}.
Leveraging UI to link privacy information does inherently focus less on non-UI components, which may lead to gaps in capturing practices that circumvent standard permission protocols~\cite{reardon201950}. Our choice is a deliberate trade-off to prioritize ease of use for all stakeholders over exhaustive documentation, fostering a more accessible approach to privacy management across different roles.

%
%
\begin{table*}[!t]

\caption{Design of PriBOM. 
One example of filled \texttt{PriBOM} of a real mobile application~\cite{LepsWorld2} is presented.}
  \label{tab:pribom_format}
\centering
\footnotesize
\resizebox{0.98\textwidth}{!}{%
\begin{tabular}{|l|l|l|}
\hline
\rowcolor{lightgray!85}
\multicolumn{3}{|c|}{\textbf{UI Widget Identifier}} \\
\hline
\rowcolor{lightgray!35}
\textbf{Data Field} & \textbf{Description} & \textbf{Example} \\ 
\hline
\rowcolor{lightgray!10}
Widget Type & Component types of the widget. & android.view.MenuItem\\ 
\hline
\rowcolor{lightgray!35}
Widget ID & A unique identifier for each UI widget component in the app. & 2131296311\\ \hline
\rowcolor{lightgray!10}
Widget Name & Names given manually for widgets to better recognize them. & action\_share\\ \hline
\rowcolor{lightgray!35}
Widget Src & Reference to source files, e.g. JPG or PNG images. & none\\ \hline
\hline
\rowcolor{lightgray!85}
\multicolumn{3}{|c|}{\textbf{Codebase and Permission}} \\
\hline
\rowcolor{lightgray!35}
\textbf{Data Field} & \textbf{Description} & \textbf{Example}\\ 
\hline
\rowcolor{lightgray!10}
Event & Specific events that the widget reacts to. & item\_selected\\ 
\hline
\rowcolor{lightgray!35}
Handler & The function or method that handles the event. & \makecell[l]{com.applovin.impl.mediation.debugger.ui.b.a:\\ boolean onOptionsItemSelected(android.view.MenuItem)}\\ 
\hline
\rowcolor{lightgray!10}
Android API Level & The minimum Android API level required by the widget. & Level 1\\ \hline
\rowcolor{lightgray!35}
Permission & The Android permissions required by the widget. & android.permission.ACCESS\_COARSE\_LOCATION\\ 
\hline
\rowcolor{lightgray!10}
Data Type & The types of data the widget collects or processes. & Location\\ 
\hline
\rowcolor{lightgray!35}
Method (Permissions) Path/Location & The path that accesses to the file requesting the permissions. & \makecell[l]{Landroid/location/LocationManager;-getLastKnownLoca\\tion-(Ljava/lang/String;)Landroid/location/Location;} \\ \hline
\hline
\rowcolor{lightgray!85}
\multicolumn{3}{|c|}{\textbf{Third-Party Library}} \\
\hline
\rowcolor{lightgray!35}
\textbf{Data Field} & \textbf{Description} & \textbf{Example}\\ 
\hline
\rowcolor{lightgray!10}
TPL Name& The package name of third-party libraries involved in the widget. & javax.inject\\ 
\hline
\rowcolor{lightgray!35}
TPL Version& The version of third-party libraries involved in the widget. & 1\\ 
\hline
\rowcolor{lightgray!10}
Latest TPL Version & The most recent version of third-party libraries available. & 1.0.0.redhat-00012\\ 
\hline
\rowcolor{lightgray!35}
TPL Publish Date (current version) & The release date of the current TPL version. & Oct 13, 2009\\ 
\hline
\rowcolor{lightgray!10}
TPL Publish Date (latest version) & The release date of the latest TPL version. & Apr 16, 2024\\ \hline
\hline
\rowcolor{lightgray!85}
\multicolumn{3}{|c|}{\textbf{Privacy Notice Disclosure}} \\
\hline
\rowcolor{lightgray!35}
\textbf{Data Field} & \textbf{Description} & \textbf{Example}\\ 
\hline
\rowcolor{lightgray!10}
Privacy Policy Description & \makecell[l]{Corresponding sections in the privacy policy related to the widget's\\ data practices.} & \makecell[l]{``with your permission we may collect your geo-location\\ information to optimize user experience, such as for\\ localization accuracy...''}\\ 
\hline
\rowcolor{lightgray!35}
Privacy Label Declaration & \makecell[l]{Disclosure of privacy practices on related data type in privacy labels.} & \makecell[l]{[Approximate Location] \\Optional: Yes; Purpose: App functionality, Analytics, Ad-\\vertising or marketing}\\ \hline

\end{tabular}
}%
\end{table*}

\vspace{-5pt}
\subsection{Design of \texttt{PriBOM}}~\label{sec_bom_design}
Figure~\ref{fig_PriBOM_flow} shows an overview of \texttt{PriBOM} usage, where it serves as a privacy information inventory indexed by UI widgets toward privacy notice generation. The proposed format is shown in Table~\ref{tab:pribom_format} with a real example. 
Recognizing that development teams may have varying requirements and constraints, \texttt{PriBOM} should not be a rigid structure but a customizable approach with modifiability that teams can adjust according to their specific needs. Within the scope of this paper, the \texttt{PriBOM} design is tailored for mobile app scenarios, owing to the ubiquitous nature and privacy concerns of mobile applications. However, the underlying concept is sufficiently robust to extend to other software development contexts.
The elaboration of data fields in \texttt{PriBOM} is listed below:



\textbf{UI Widget Identifier Section.} As discussed in Section~\ref{sec_motv}, UI widgets are the pivot to synergistically connect different roles of developers on privacy-related communication and collaboration. Therefore, we set the granularity of \texttt{PriBOM} at the UI widget level and include UI identifier in \texttt{PriBOM} to achieve sufficient identification.


\begin{itemize} [leftmargin=*, noitemsep, topsep=3pt]
    \item \textbf{[Widget ID]}  \emph{Widget ID} is the unique identifier for each widget and serves as fundamental information for referencing and mapping widgets to other data fields in \texttt{PriBOM}. 
    \item \textbf{[Widget Type]} \emph{Widget Type} records the functional type of a widget,  e.g. ``android.widget.ImageView'' for image display, enabling developers to quickly form a preliminary understanding of the widget and achieve more fine-grained categorization and easier management. 
    \item \textbf{[Widget Name]} \emph{Widget Name} is a readable common name given by developers to better recognize them during implementation and maintenance. 
    \item \textbf{[Widget Scr]} \emph{Widget Scr} is the reference to the widget's source files. For example, a functional button may be associated with a PNG icon image for appearance. 
\end{itemize}

\textbf{Codebase and Permission Section.} This section focuses on the intricate relationship between the codebase specifics and the widgets. It covers data fields like the specific events that a widget reacts to, the methods of handling these events, the Android API levels, and the permissions required by each widget. Documenting these details aids developers in ensuring that the application adheres to best practices in privacy, facilitating an organized approach to managing privacy effectively.
Our pre-fill of \texttt{PriBOM} utilizes static analysis to fill out this section, which is discussed in detail in Section ~\ref{source_code_ana_module}.

\begin{itemize} [leftmargin=*, noitemsep, topsep=3pt]
    \item \textbf{[Event]} \emph{Events} represent the triggers or user actions to which the widget responds. For example, ``click'' means the widget is clickable. This data field helps to understand the interactions leading to possible data processing.
    \item \textbf{[Handler]} \emph{Handler} refers to the method that responds to the \emph{event}, which identifies the specific code block handling the \emph{event's} response. For example, the \emph{handler} of \emph{event} ``click'' can be ``void onClick(android.view.View)''. 
    \item \textbf{[Android API Level]} This field records the minimum required Android API level the widget can operate on. \emph{Android API Level} is considered as key information to ensure compatibility with various Android OS versions.
    \item \textbf{[Permission]} Android utilizes app permission mechanism~\cite{PermissionsonAndroid} to protect access to both restricted data, e.g. Contacts, and restricted actions, e.g. taking pictures. More specifically, an app must require runtime permissions before it obtains access to additional data or performs actions that may affect the system, other apps or devices. Such requests may be involved in the widget's behind-the-scene behaviors~\cite{xi2019deepintent}. For example, an elliptical coordinate icon in a food delivery app is likely to be accompanied by a request to grant geographic location permissions. Such requests are raised by the Android API calls in the widget callbacks. 
    \item \textbf{[Method (Permissions) Location]} This field documents where in the codebase permissions are handled or required, e.g., the file path or location that accesses the permissions related to the widget.
    \item \textbf{[Data Type]} This field records the specific data type required by corresponding Android Permission, which is also regarded as the data that the widget collects or processes. Following the grouping strategy in previous researches~\cite{mcconkey2023runtime, rahman2022permpress} and Official Android API Documentation~\cite{Manifestpermission}, we map the Android Permissions to specific data types. For example, ``READ\_CONTACTS'' and ``WRITE\_CONTACTS'' are categorized into \emph{data type} ``Contacts''. A more detailed discussion is in Section~\ref{source_code_ana_module}.

\end{itemize}

\textbf{Third-Party Library Section.} This section is dedicated to managing and documenting the use of TPLs associated with the widget, which play an important role in app functionality and data practices. It records the basic info of used TPLs. This section is crucial for maintaining up-to-date TPL usage and mitigating risks associated with outdated components. 

\begin{itemize} [leftmargin = *]
    
    \item \textbf{[TPL Information]} Third-party libraries (TPL) pervasively serve as reusable functional components for mobile apps to improve development efficiency~\cite{wu2023libscan}. For example, the in-app payment service of a shopping app may be supported by TPLs. \texttt{PriBOM}, therefore, records the baseline information of TPLs relevant to the widget, including \emph{name, version, latest version, publish date (current version), and publish date (latest version)}. \texttt{PriBOM} documents the publish date of the currently used and most up-to-date version of TPLs to enable assessing the currency of the version in use and informing about potential updates needed for security or functionality enhancements.  

\end{itemize}

\textbf{Privacy Notice Disclosure Section.} 
In application development, a significant portion of effort is 
 often allocated to ongoing updates rather than the outset. This principle also applies to privacy notices, given the evolving nature of legal and regulatory requirements. Consequently, \texttt{PriBOM} effectively manages two common forms of privacy notices, e.g. privacy policies and privacy labels, by linking widget components directly to their respective disclosures, 
 ensuring continuous alignment between app data practices and legal requirements throughout the development lifecycle.
 


\begin{itemize} [leftmargin=*, noitemsep, topsep=3pt]
    \item \textbf{[Privacy Policy Description]} This field refers to the disclosures in the app's privacy policy related to the widget’s data practices. By recording relevant descriptions of the privacy behaviors, \texttt{PriBOM} links the technical implementation of the widget with the publicly stated privacy practices, offering a window for privacy compliance checking.
    \item \textbf{[Privacy Label Declaration]} Before publishing apps on markets such as Google Play, developers must declare their app's privacy practices, e.g. data collection and handling, in a privacy nutrition label form. However, developers often struggle to create authoritative and accurate privacy labels~\cite{li2022understanding, balebako2014improving, balebako2014privacy}. 
    Including privacy labels' data practice declarations that match the data type in \texttt{PriBOM} can support developers and legal teams in aligning privacy disclosure with actual app behaviors.


    
\end{itemize}


\begin{figure*}[t]
  \centering
  \includegraphics[width=.8\linewidth]{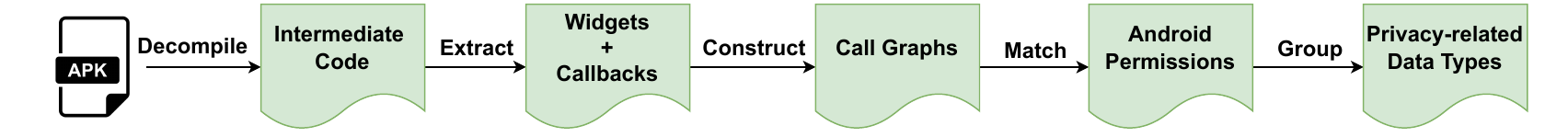}

  \caption{The pipeline of our static analysis module.}
  \label{fig_stat_pipe}
\vspace{-5pt}
\end{figure*}

\subsection{Benifits of \texttt{PriBOM}}
Presenting the design of \texttt{PriBOM} reveals several unique benefits of integrating \texttt{PriBOM} into DevOps, responding to the challenges identified in the formative study:

\noindent \textbf{[Benefit-1] Transparency.} Lacking team-level and organizational support is considered a challenge faced by developers~\cite{li2018coconut, li2022understanding}. 
Therefore, to address the privacy ambiguity caused by the lack of team support and assistant tools, the necessity of a recording and communication tool focused on privacy information has become more prominent. \texttt{PriBOM} can serve as a central communication tool, bridging the informational gap between developers of different roles from various teams. By providing a unified view of privacy practices at the widget level, \texttt{PriBOM} can help all developers have a consistent understanding of potential privacy implications, fostering informed decision-making and cohesive development strategies. Through \texttt{PriBOM}, transparency can be achieved as every practitioner, regardless of their role, gains clear insights into privacy aspects.


\noindent \textbf{[Benefit-2] Privacy Traceability and Trackability.} \texttt{PriBOM} connects the frontend UI widget, the backend privacy practices in source code, and the app's privacy notices, helping developers gain traceability and trackability on this \emph{privacy chain}. Similar to terminology in software supply chain~\cite{kelepouris2007rfid}, both terms in the context of \texttt{PriBOM} with ‘where-from’ and ‘where-used’~\cite{petroff1991framework} directions respectively are as follows:

\begin{itemize} [leftmargin=*, noitemsep, topsep=3pt]
\item \emph{Tracing.} Given any point in this privacy chain, \texttt{PriBOM} enables developers to work backward to query privacy-relevant information. For example, if a user reports privacy issues in interacting with certain UI widgets, then the development team can easily track back and locate the problematic code files (Currently, a privacy breach takes an average of 287 days to completely resolve~\cite{simplelegal}).

\item \emph{Tracking}. \texttt{PriBOM} enables developers to work forwards to find corresponding UI widgets and privacy notice content based on certain privacy practices. 
For example, if a code change leads to a change in privacy practices, the development team can find all relative privacy notice entries and update them.
Such trackability greatly contributes to the privacy maintenance after the initial software development.
\end{itemize}

\noindent \textbf{[Benefit-3] Collaboration.} 
Crafting privacy notices is a challenging task that requires both the knowledge of app features and legal requirements. 
Individual developers may misunderstand the app’s data practices, and misinterpretations of the relevant terminology during the creation of privacy notices may also exist.
Thus, generating accurate privacy notices requires the close collaboration of the whole development team instead of relying on one or several individuals.
Enabling by the \texttt{PriBOM}, roles like in-house legal teams or legal services providers can work closely with the developers in the task of creating privacy notices.
Additionally, the \texttt{PriBOM} acts as a comprehensive inventory encapsulating a detailed record of data handling practices, providing clear references for internal privacy inquiries about all roles.


\section{Pre-fill of \texttt{PriBOM}}
\label{sec_imple}

In this section, we present a pre-fill implementation of \texttt{PriBOM}, and describe details regarding the pipeline. Our pre-fill implementation is built upon existing tools~\cite{rountev2014static, yang2015static, yang2018static, Androguard, wu2023libscan, pan2023seeprivacy}.
Notably, although our design tailors to Android apps, the concept of \texttt{PriBOM} can be easily adapted on iOS apps with corresponding modifications.

\subsection{Static Analysis}
\label{source_code_ana_module}



Following Android analysis approaches in previous icon-behavior researches~\cite{xi2019deepintent, zhao2021icon2code, malviya2023fine}, our static analysis-based pre-fill implementation consists of four steps: (1) Widget and Callback Method Extraction; (2) Call Graph Construction; (3) Android Permission Extraction; and (4) Permission-Data Mapping. Figure~\ref{fig_stat_pipe} provides an overview of the pipeline.


\textbf{Widget and Callback Method Extraction}. In the first step, the implementation pipeline takes the Android application package (APK) file as input, and leverages JADX~\cite{jadx} to disassemble it. The purpose of this step is to extract widget components, associated sources such as icons, and corresponding callback methods that respond to UI interactions. 
Icons (such as PNG or JPG files) are commonly bound with UI widgets in the layout configuration file. For example, XML attribute \textit{android: background} is used to set a background image for a widget component. Therefore, we parse the layout configuration files based on a summary list~\cite{zhao2021icon2code, avdiienko2017detecting} of XML attributes used to bind icons to extract the icon sources associated with the widgets. 
For UI interactions like clicks, we extract events and their corresponding callback methods, such as \textit{onClick}. Widgets in Android apps may bind to callback methods either through XML attributes or programmatically through API calls like \textit{findViewByID}, which links to a unique widget ID. To analyze these bindings comprehensively, we employ GATOR~\cite{rountev2014static, yang2015static, yang2018static}, an Android static analysis tool based on Soot~\cite{lam2011soot}, to extract widgets and callback methods.


\textbf{Call Graph Construction.} 
We employ AndroGuard~\cite{Androguard}, a popular Android static analysis tool, to construct a call graph for each identified callback methods. These graphs elucidate the invocation relationships between methods, with nodes representing methods and edges indicating their interactions. With a callback method as the entry point, each graph not only tracks direct API invocations but also the cascade of method calls they may initiate. As the call graph of a callback method can be considered as a subgraph of the call graph of the entire app~\cite{xi2019deepintent}, our pre-fill pipeline resorts to AndroGuard~\cite{Androguard} to construct a complete call graph of the app, and then extract the call graph for each identified callback method. Ultimately, mappings between the widget component and its reachable methods can be obtained.


\textbf{Android Permission Extraction.} This step maps each widget component to its associated potential Android permission requests. Our pre-fill pipeline traverses the call graph of the widget's callback method to extract all the API calls, and leverages AndroGuard~\cite{Androguard} to obtain a mapping from APIs to Android permissions and determine which permissions are requested. This step associates widget components with their corresponding Android permission uses.


\textbf{Permission-Data Mapping.} Our pre-fill pipeline establishes associations between widget components and interaction-driven data practices by mapping Android permissions to relevant data types. Following the approach in prior works~\cite{rahman2022permpress, mcconkey2023runtime}, we group the dangerous Android permissions into ten different data type categories. For example, a widget associated with Android permission \textit{ACCESS\_COARSE\_LOCATION} is considered to be related to data type \textit{Location}, e.g., the user interaction with this widget leads to the application's data practice of accessing the user's location information. Dangerous Android permissions refer to permissions with a protection level of \textit{dangerous} in the Official Android API Documentation~\cite{Manifestpermission}.




\textbf{TPL Detection.} Third-party libraries (TPLs) are constitutional in mobile app development. However, insufficient understanding of TPLs used may lead to developers' non-comprehensive understanding of the privacy behaviors of the application~\cite{khandelwal2023unpacking, li2022understanding, li2018coconut, li2021developers, balebako2014privacy}. We adapt the existing state-of-the-art TPL detector, LibScan~\cite{wu2023libscan}, to extract the TPLs and version information based on code features and sophisticated similarity analysis. 
It establishes pairwise class correspondences between app and TPL classes, computes confidence scores, and determines TPL presence.

Notably, our pre-fill pipeline is highly modularized, which means each individual module, like the TPL detector, can be easily replaced by more powerful ones. Therefore, we are not overly fixated on performance but instead provide developers and companies with the freedom to modify module implementation choices through coupling in real-world scenarios.

\subsection{Privacy Notice Analysis}

Privacy notice generation involves not just initial creation but also continual updates and maintenance. A common scenario is maintaining consistency between existing privacy notices, evolving software, and updating regulations.
To enhance the alignment between privacy prompts and actual data practices in applications, \texttt{PriBOM} includes the Privacy Notice Disclosure section. The pre-fill implementation of this section analyzes and processes two common forms of privacy notices, privacy policy and privacy label, and matches them with the corresponding data types documented in \texttt{PriBOM}.


\textbf{Privacy Policy Segmentation.} The objective of this process is to obtain privacy policy segments related to certain data types. For example, segments of data type \textit{Location} will include all descriptions related to \textit{location} in the privacy policy, such as whether it will be collected and the purpose of such collection. We adopt the multi-level privacy policies processing strategy in ~\cite{pan2023seeprivacy, xie2022scrutinizing} to perform segment extraction, which takes the app's privacy policy as input and extract sentences related to each data type through paragraph-level headings classification and sentence-level keyword searching and phrase similarity calculation.
We employ prior work~\cite{pan2023seeprivacy, xie2022scrutinizing} for this privacy policy extraction and segmentation process. 


\textbf{Privacy Label Processing.} This workflow extracts privacy practice disclosures from the application's privacy label, e.g. the DSS~\cite{Datasafetysection} in Google Play. The privacy label information in the DSS includes the collected data type, the purpose of collection, and whether this collection is optional. 
Specifically, we record the existence and collection purpose of the data type, as well as whether the granting of this data type is optional for application users. We then map the data types with the data types extracted through the static analysis pipeline discussed in section~\ref{source_code_ana_module}. An example of \texttt{PriBOM} is presented in Table~\ref{tab:pribom_format} of a real mobile application~\cite{LepsWorld2} for the UI widget whose ID is \textit{``2131296311''} and the type is \textit{android.view.MenuItem}.
From the PriBOM, developers can easily obtain that this UI widget is related to permission request, \textit{android.permission.ACCESS\_COARSE\_LOCATION}, for coarse positioning and indicates its corresponding data type, \textit{Location}.
In addition, from the Privacy Notice Disclosure section, developers can notice the disclosure in the privacy policy about \textit{Coarse Location}, \textit{``...may collect your geo-location information to optimize user experience...''}; and corresponding privacy label \textit{Approximate Location} information, which is \textit{optional} for data collection and \textit{App functionality, Analytics, Advertising or marketing} as its purposes. 

\section{Human Evaluation}
\label{sec_hum_eval}

To explore \texttt{PriBOM}'s usefulness and prompt further requirements, we conducted a totally anonymous online survey to examine the perspectives of people in various software development roles on \texttt{PriBOM}.
Ethical approval for this research was secured from our institution’s Institutional Review Board (IRB). For more information, please refer to Section~\ref{Ethics}.



\subsection{User Study Design}
\label{survey_design}

The user study aims to evaluate the perceived usefulness of \texttt{PriBOM} and prompt insights and further requirements to adapt to specific needs in the real world. 
To ensure participants fully understand our \texttt{PriBOM} design, we introduce each section of \texttt{PriBOM} individually and provide an example, as shown in Table~\ref{tab:pribom_format}. 
In addition, to make sure participants are properly prompted, we deliberately put the data field and descriptions on the side for quick reference. 

For the questionnaire, we developed our survey questions based on the specific section design of \texttt{PriBOM}
and perspectives from related studies and guidelines~\cite{balebako2014privacy, weir2020needs, xia2023empirical, kitchenham2008personal}. After iterative refinement and pilot studies with 10 participants in total, our final survey contains 26 statements, with which participants rate their agreement levels on a 5-point Likert scale ranging from 1 (strongly disagree) to 5 (strongly agree).
Table~\ref{tab_statements} enumerates the statements. 
The questions covers six perspectives: (1) General Design of \texttt{PriBOM}, (2) Widget Identifier, (3) Codebase and Permission, (4) TPL, (5) Privacy Notice Disclosure, and (6) Usability and Practicality. 
Drawing on the pilot study results, we also include four free-text open-ended questions to seek deeper perspectives from participants. The open-ended questions are listed in Table~\ref{tab_open_question}.

We target participants with working experience in software development, including various roles in the development team, such as creative design, software engineering, legal, and product management. The responses of pilot study are excluded from the final results.

%
\begin{table}[t]
\centering
\caption{Participants demographics. Each answer is labeled with the count of participant(s) that select it. ``N.Am.'' stands for ``North America''.}
\label{tab_demo}
\resizebox{0.8\linewidth}{!}{%
\begin{tabular}{llll}
\toprule
\textbf{Role}   & \textbf{Team Size} & \textbf{Gender} & \textbf{Continent}\\
\midrule
Junior developer (59)  & \textless 10 (89)& Male (94) & Europe (85)\\
Senior developer (19) &10 - 20 (29) & Female (55) & Africa (45)\\
Project manager (21)   & 20 - 50 (19) & Unknown (1) & Asia (12)\\
UI designer (22)   & \textgreater 50 (13) & - & N.Am. (5)\\
Legal team (13)   & - & - & Oceania (3)\\
Others (16)   & - & - & -\\
\bottomrule
\end{tabular}
}%
\vspace{-10pt}
\end{table}

%

%
\begin{table*}[t]
\centering
\caption{Our survey results of participants' agreement on statements related to PriBOM. ``Ave.'' stands for the average Score, and a higher score indicates stronger agreement. ``Distr.'' denotes the distribution of the responses (from left to right: strongly disagree to strongly agree).}
\vspace{-10pt}
\label{tab_statements}
\resizebox{1.0\textwidth}{!}{%
\begin{tabular}{l|l|c|c}
\toprule

\textbf{No.} & \textbf{Statement} & \textbf{Ave.} & \textbf{Distr.} \\

\midrule
\multicolumn{4}{c}{\textbf{General Design of PriBOM}} \\
\midrule

$\textit{S}_1$  &\makecell[l]{\textbf{[Intuitiveness]} The data fields in PriBOM appear logically structured and are understandable.} & 3.91 & \raisebox{-0.32\totalheight}{
\includegraphics[width=0.04\textwidth]{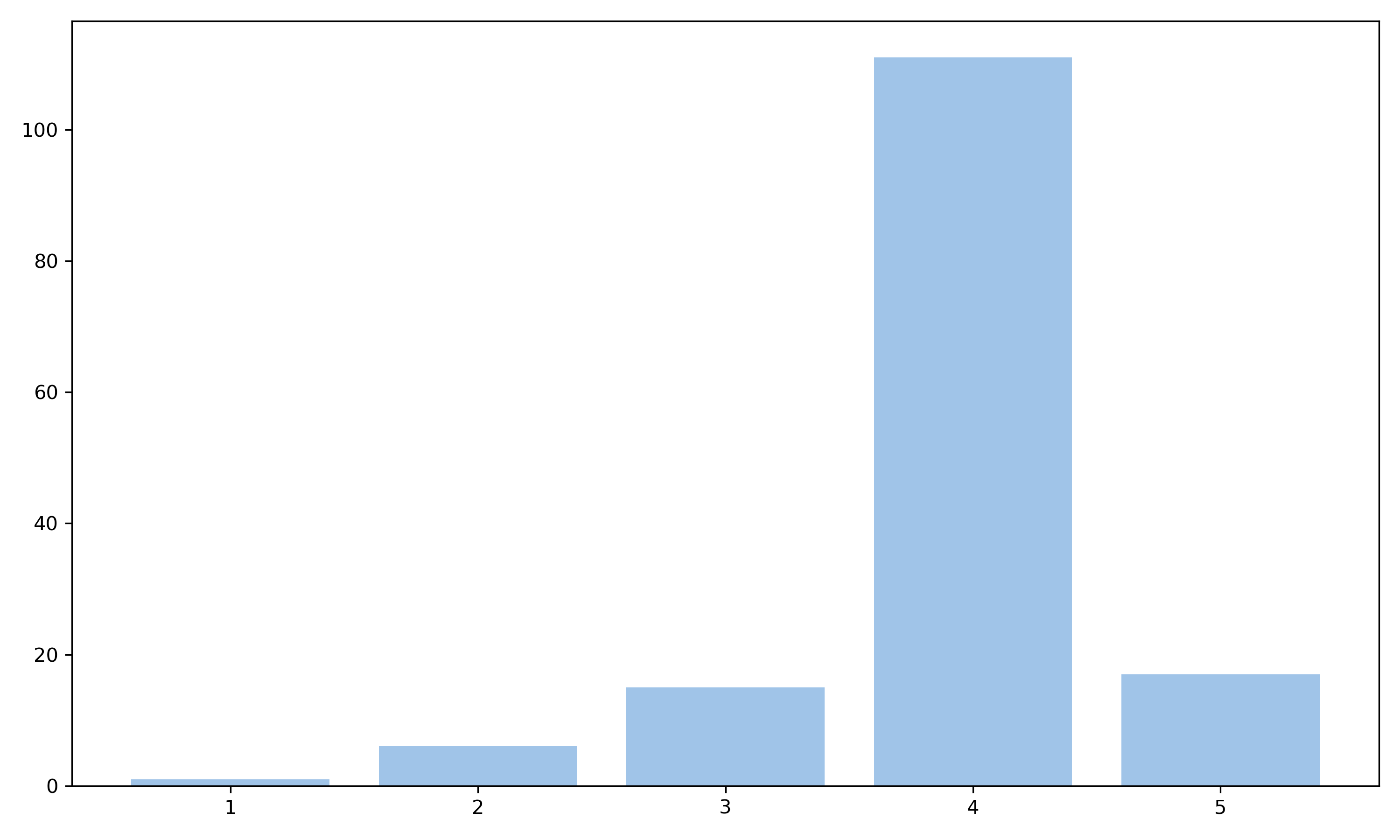}
}

\\

$\textit{S}_2$  &\makecell[l]{\textbf{[Format]} The layout and format of PriBOM are intuitive for developers with varying levels of experience.} & 3.83 & \raisebox{-0.32\totalheight}{\includegraphics[width=0.04\textwidth]{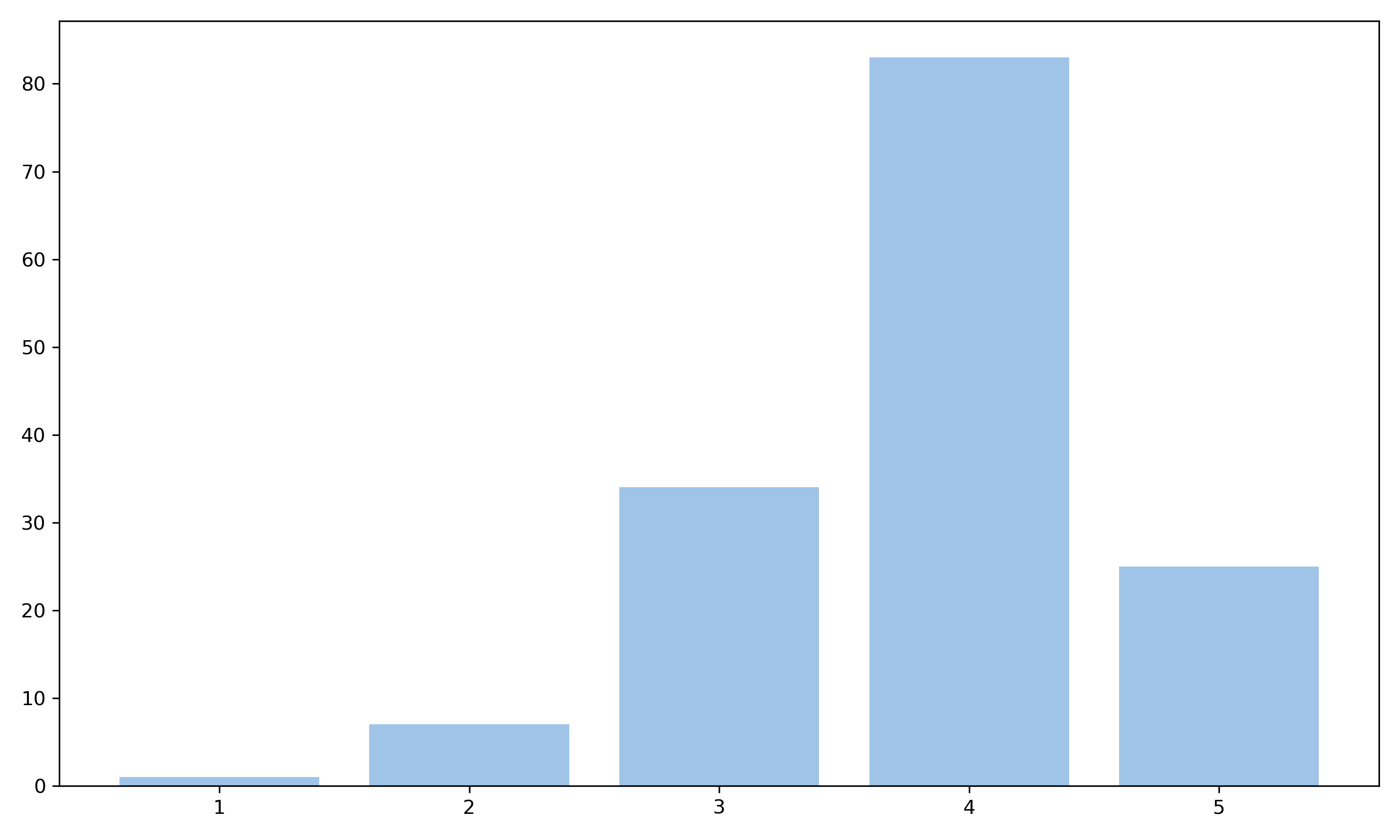}} \\
$\textit{S}_3$  &\makecell[l]{\textbf{[Relevance]} Information in the PriBOM is essential and contributes to privacy management.} & 4.02 & \raisebox{-0.32\totalheight}{\includegraphics[width=0.04\textwidth]{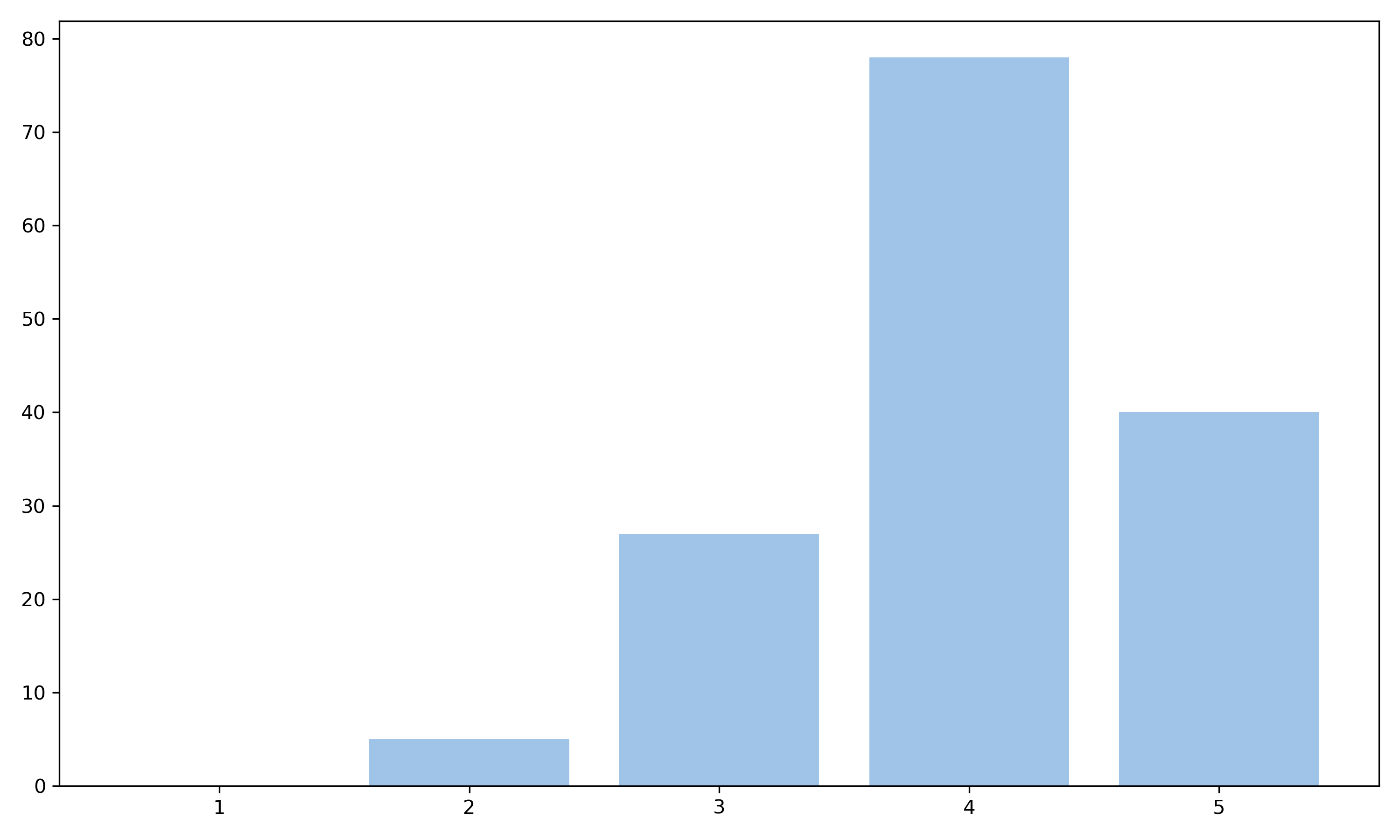}} \\
\midrule

\multicolumn{4}{c}{\textbf{Design of the Widget Identifier Section of PriBOM}} \\
\midrule
$\textit{S}_4$  &\makecell[l]{\textbf{[Identification Precision]} PriBOM's design for widget identification is precise enough to pinpoint privacy issues to a specific UI component.} & 3.69 & \raisebox{-0.32\totalheight}{\includegraphics[width=0.04\textwidth]{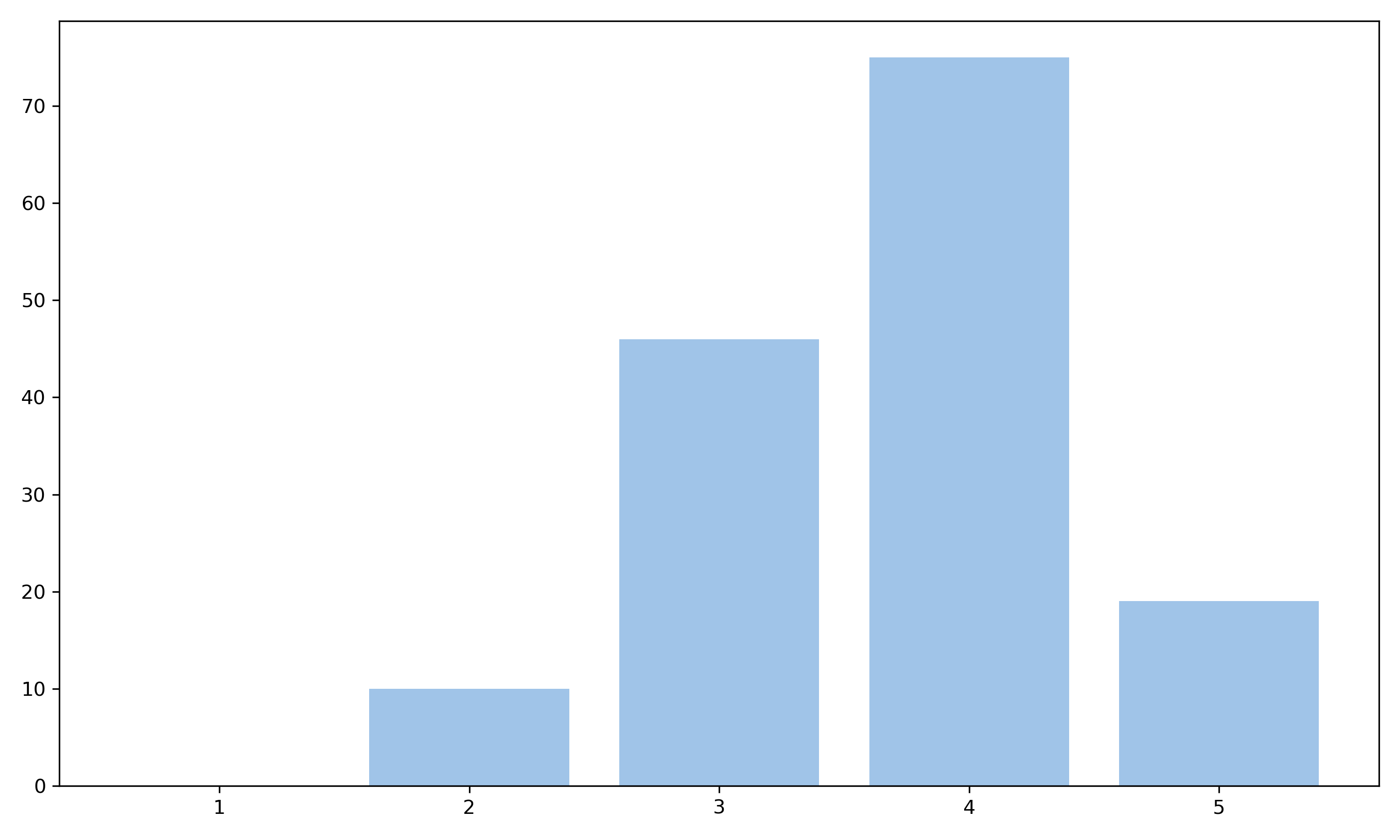}} \\
$\textit{S}_5$  &\makecell[l]{\textbf{[Clarity]} The widget name and type in the PriBOM is helpful in providing clear and common terminology for team members.} & 4.05 & \raisebox{-0.32\totalheight}{\includegraphics[width=0.04\textwidth]{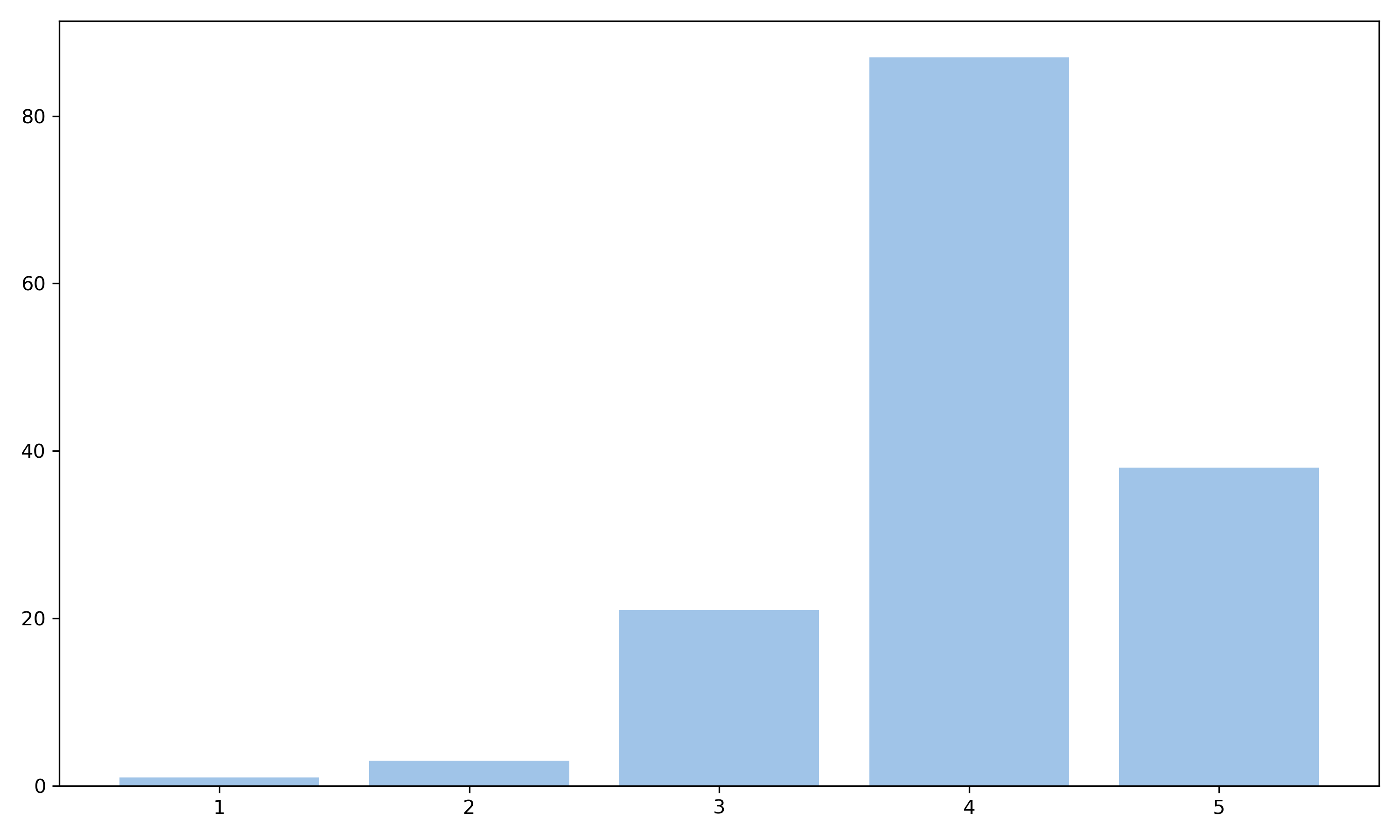}} \\
\midrule

\multicolumn{4}{c}{\textbf{Design of the Codebase and Permission Section of the PriBOM}} \\
\midrule
$\textit{S}_6$  &\makecell[l]{\textbf{[Codebase Accessibility]} The path/location fields in PriBOM intended for codebase access promote better manageability.} & 3.76 & \raisebox{-0.32\totalheight}{\includegraphics[width=0.04\textwidth]{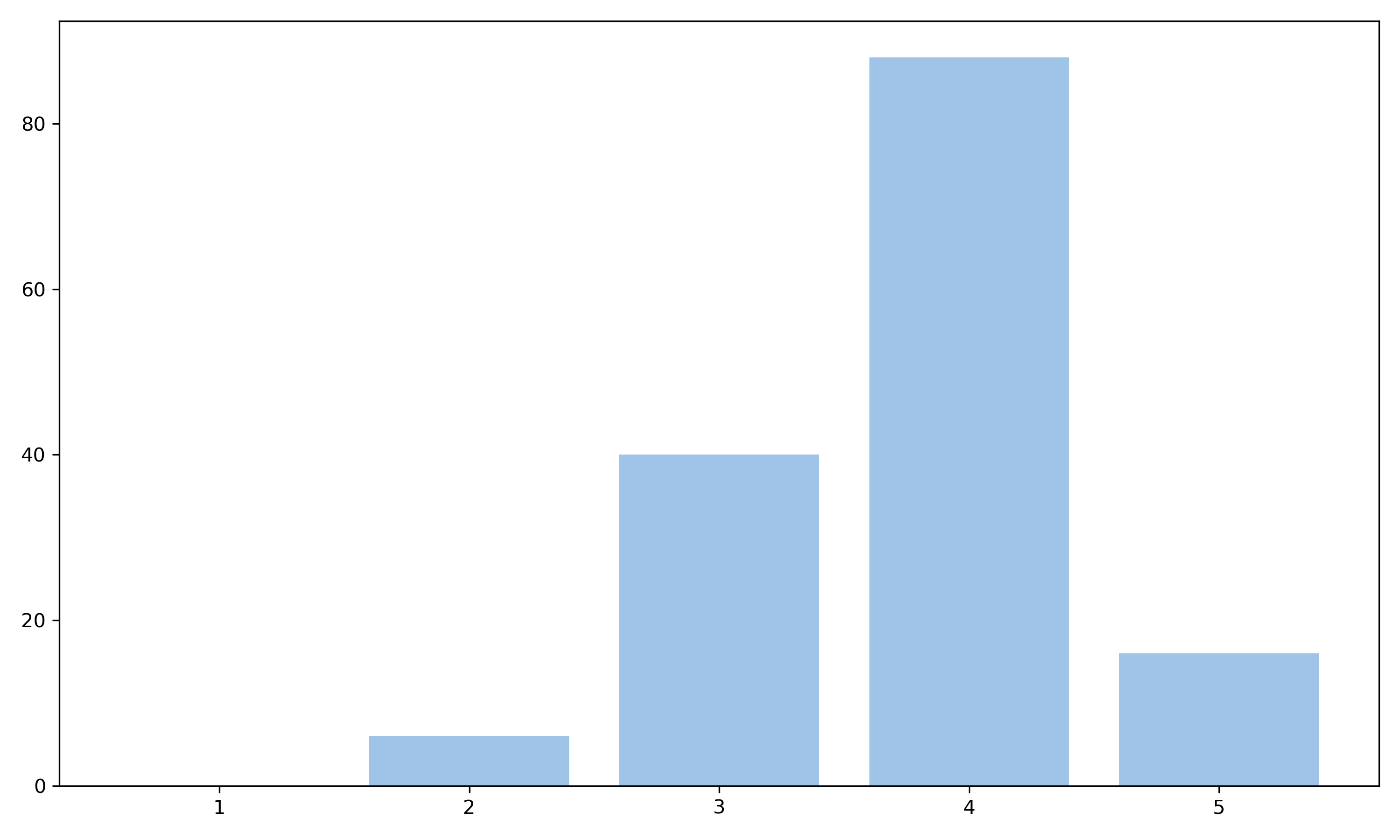}} \\
$\textit{S}_7$  &\makecell[l]{\textbf{[Event \& Handler]} The `Event' and `Handler' data field in PriBOM aid in tracing data flows in response to user events.} & 3.92 & \raisebox{-0.32\totalheight}{\includegraphics[width=0.04\textwidth]{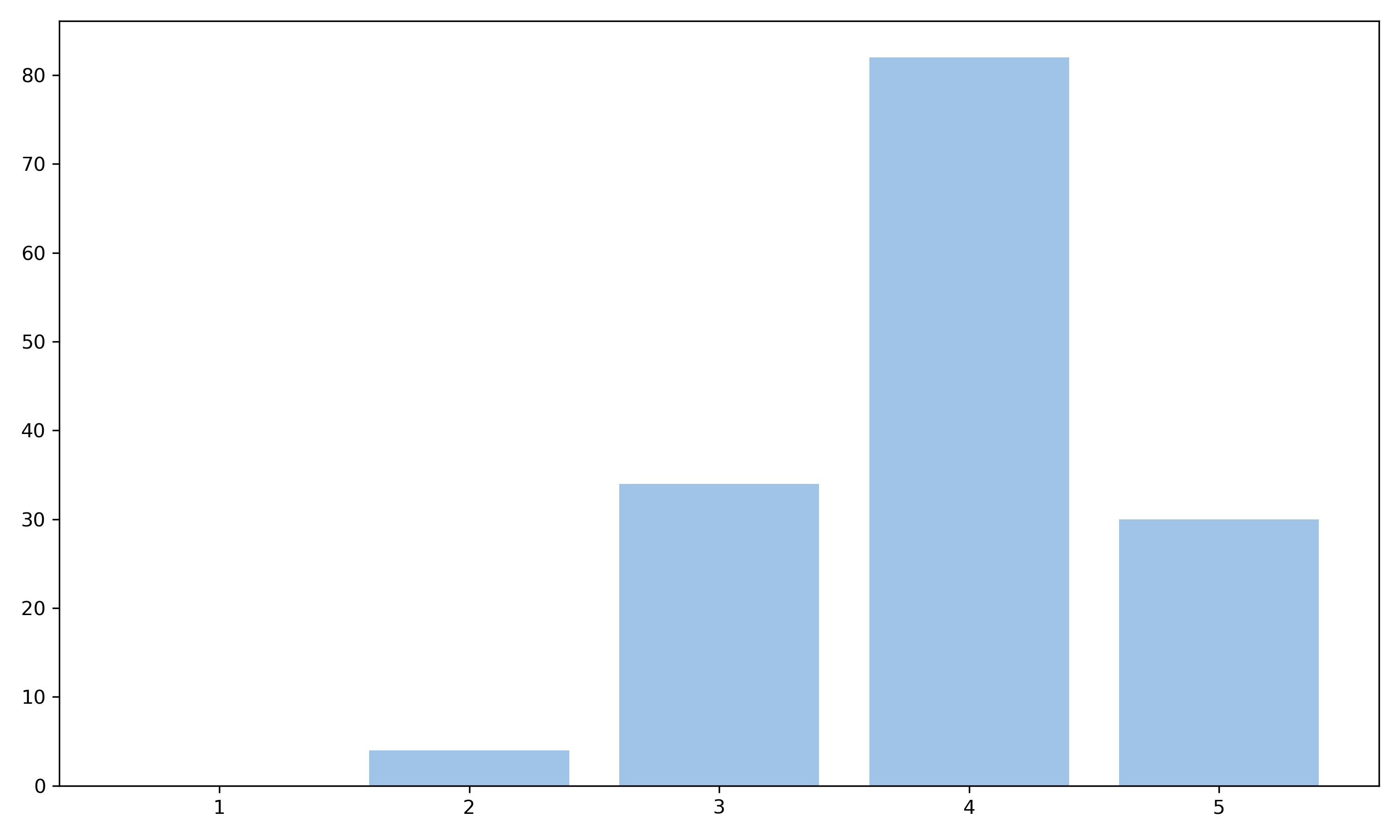}} \\
$\textit{S}_8$  &\makecell[l]{\textbf{[Permission]} Listing ‘Permission’ requirements in PriBOM may increase transparency regarding data access needs and permission requests.} & 4.09 & \raisebox{-0.32\totalheight}{\includegraphics[width=0.04\textwidth]{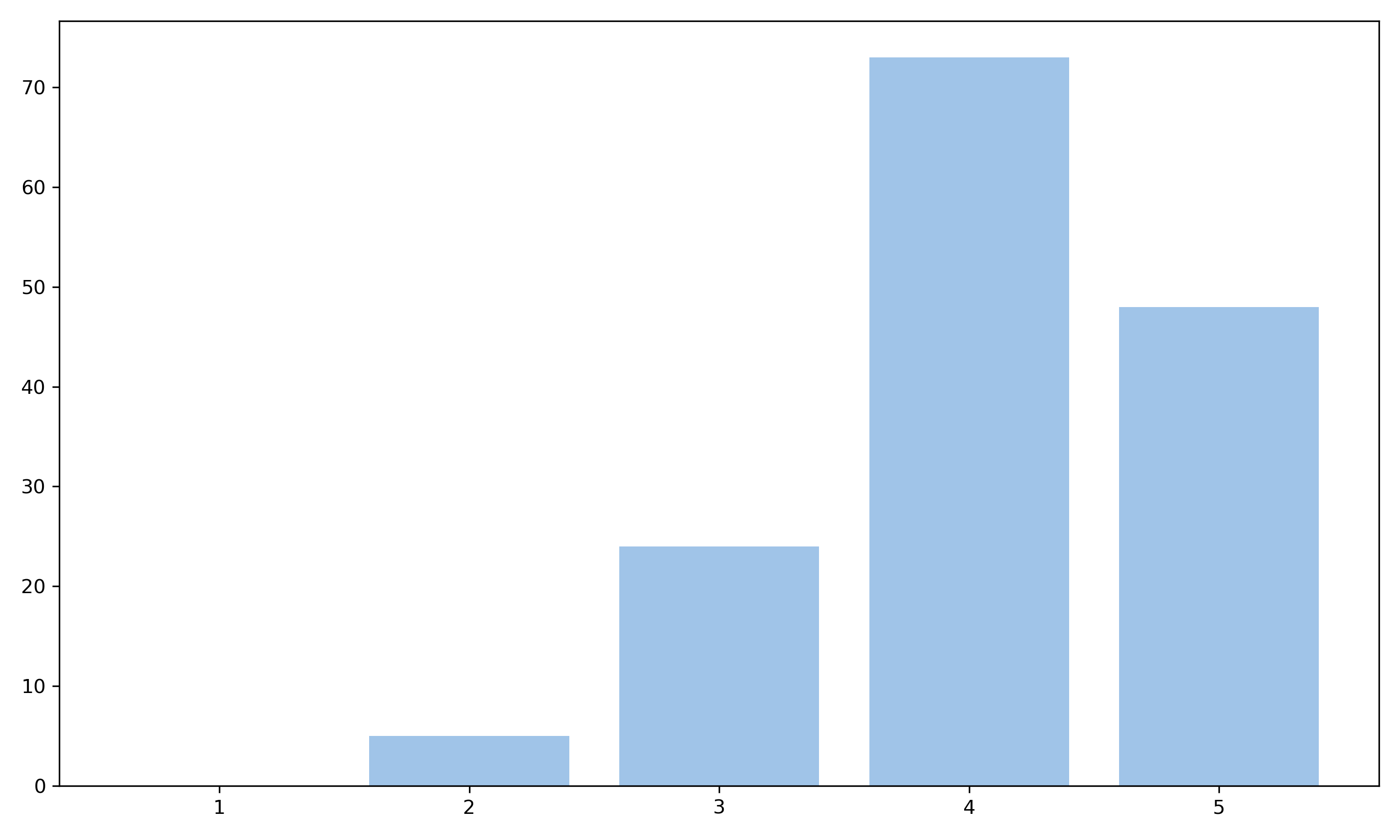}} \\

$\textit{S}_{9}$  &\makecell[l]{\textbf{[Data Type]} Including data types and associated permissions in PriBOM is instrumental for providing privacy information.} & 4.01 & \raisebox{-0.32\totalheight}{\includegraphics[width=0.04\textwidth]{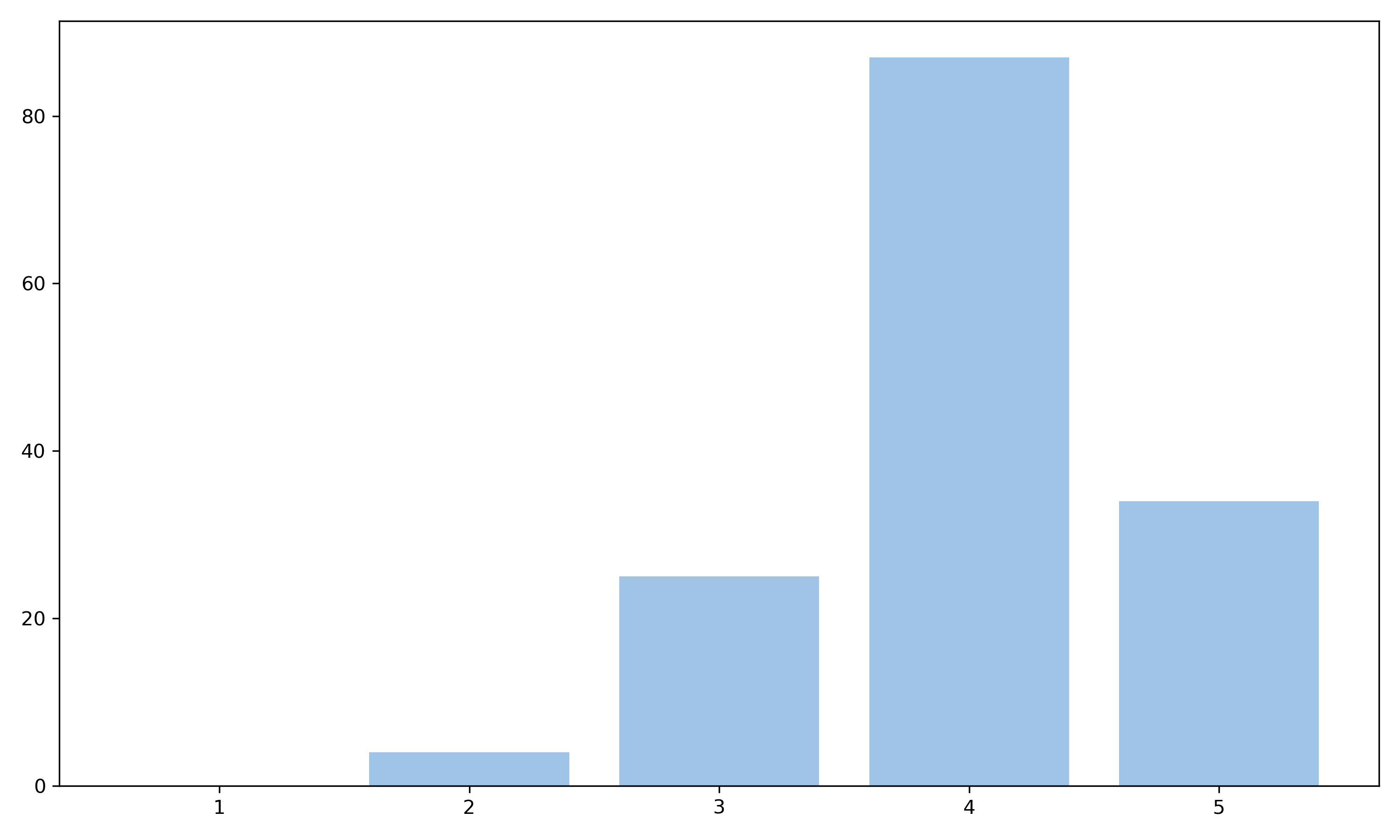}} \\

$\textit{S}_{10}$  &\makecell[l]{\textbf{[API-Level Awareness]} The inclusion of specific Android API levels in PriBOM is relevant and necessary for privacy documentation.} & 3.73 & \raisebox{-0.32\totalheight}{\includegraphics[width=0.04\textwidth]{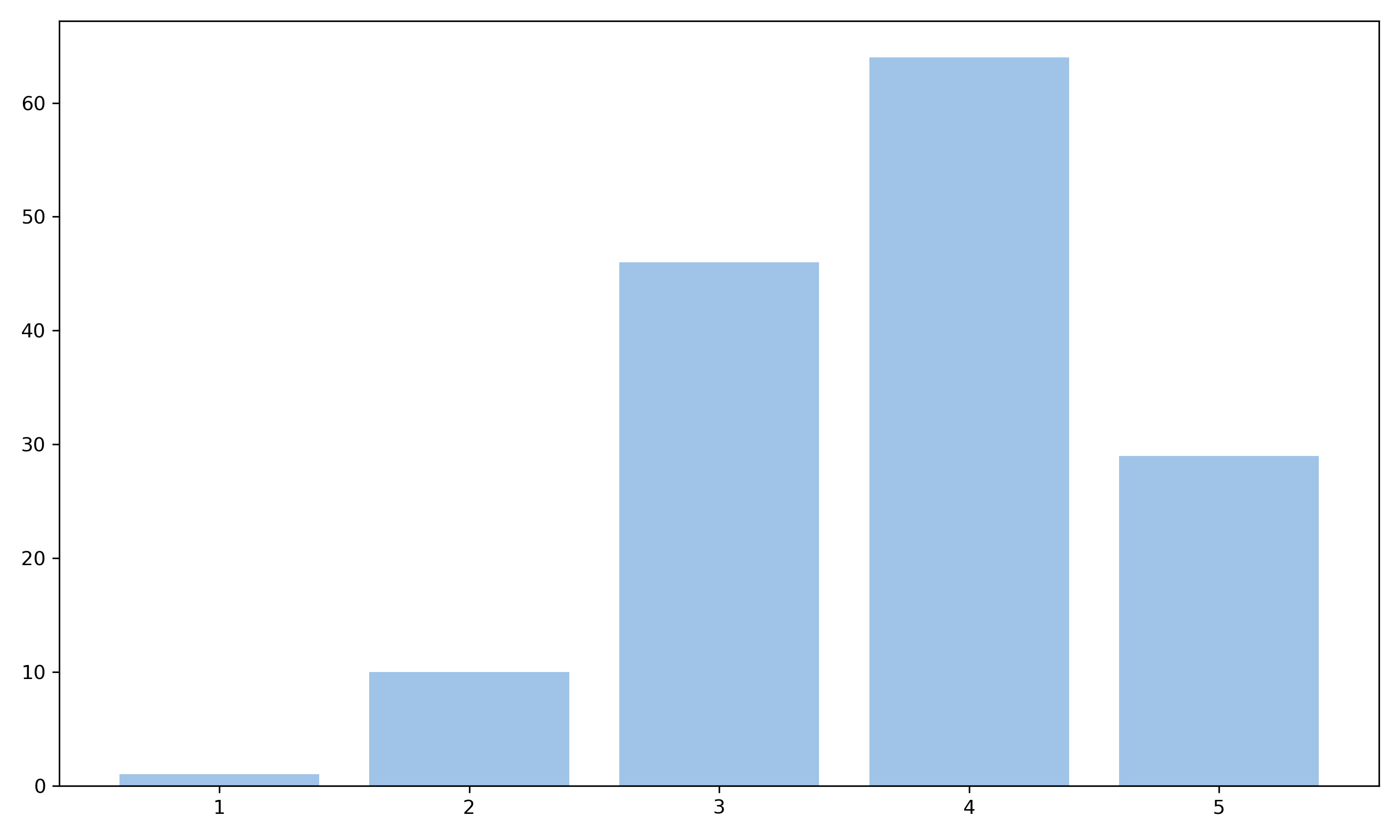}} \\

\midrule

\multicolumn{4}{c}{\textbf{Design of the TPL Section of PriBOM}} \\
\midrule

$\textit{S}_{11}$  &\makecell[l]{\textbf{[Management]} Including third-party library information in PriBOM may support a more thorough privacy review.} & 3.85 & \raisebox{-0.32\totalheight}{\includegraphics[width=0.04\textwidth]{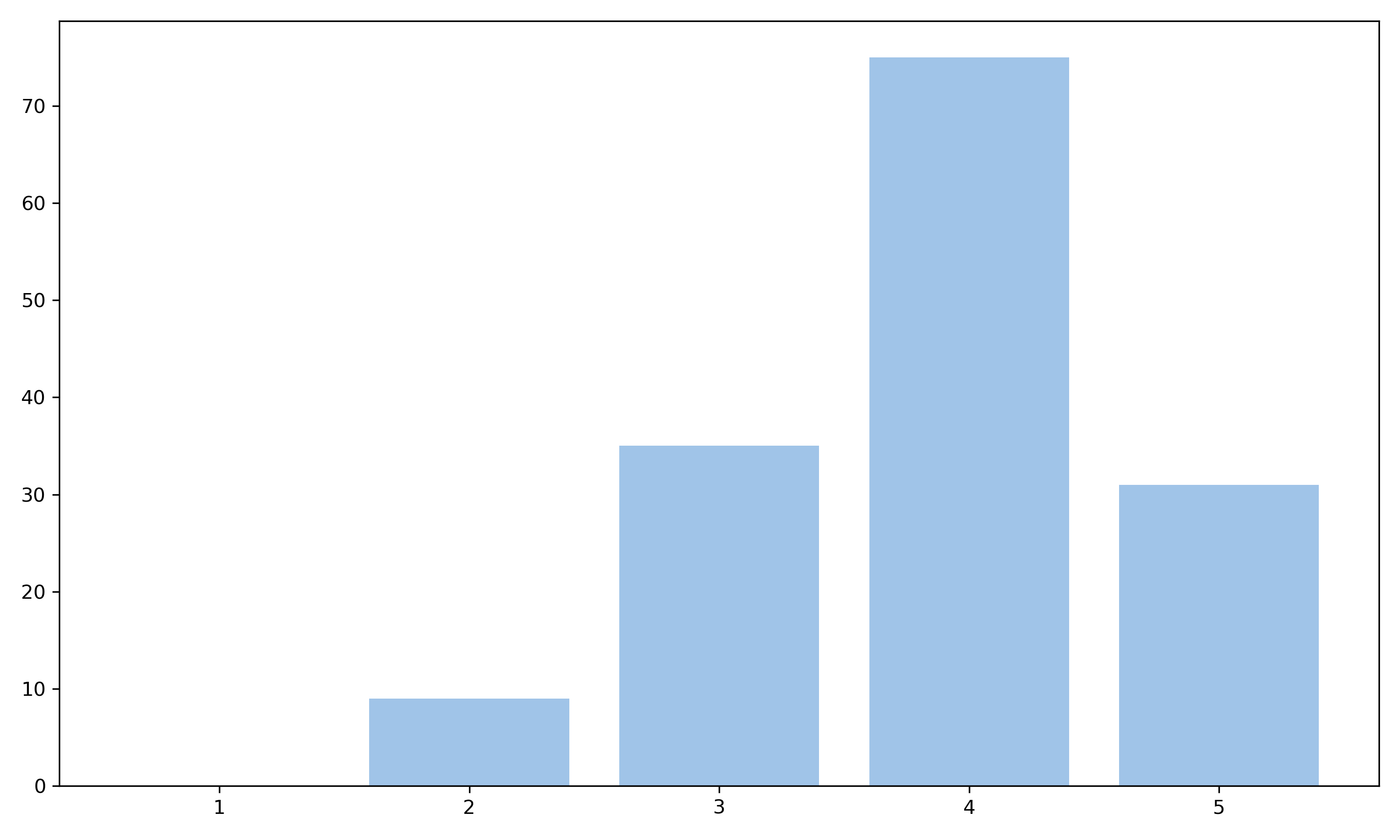}} \\
$\textit{S}_{12}$  &\makecell[l]{\textbf{[Discrepancy]} Documenting TPL versions may help identify discrepancies of privacy practices between different versions.} & 3.97 & \raisebox{-0.32\totalheight}{\includegraphics[width=0.04\textwidth]{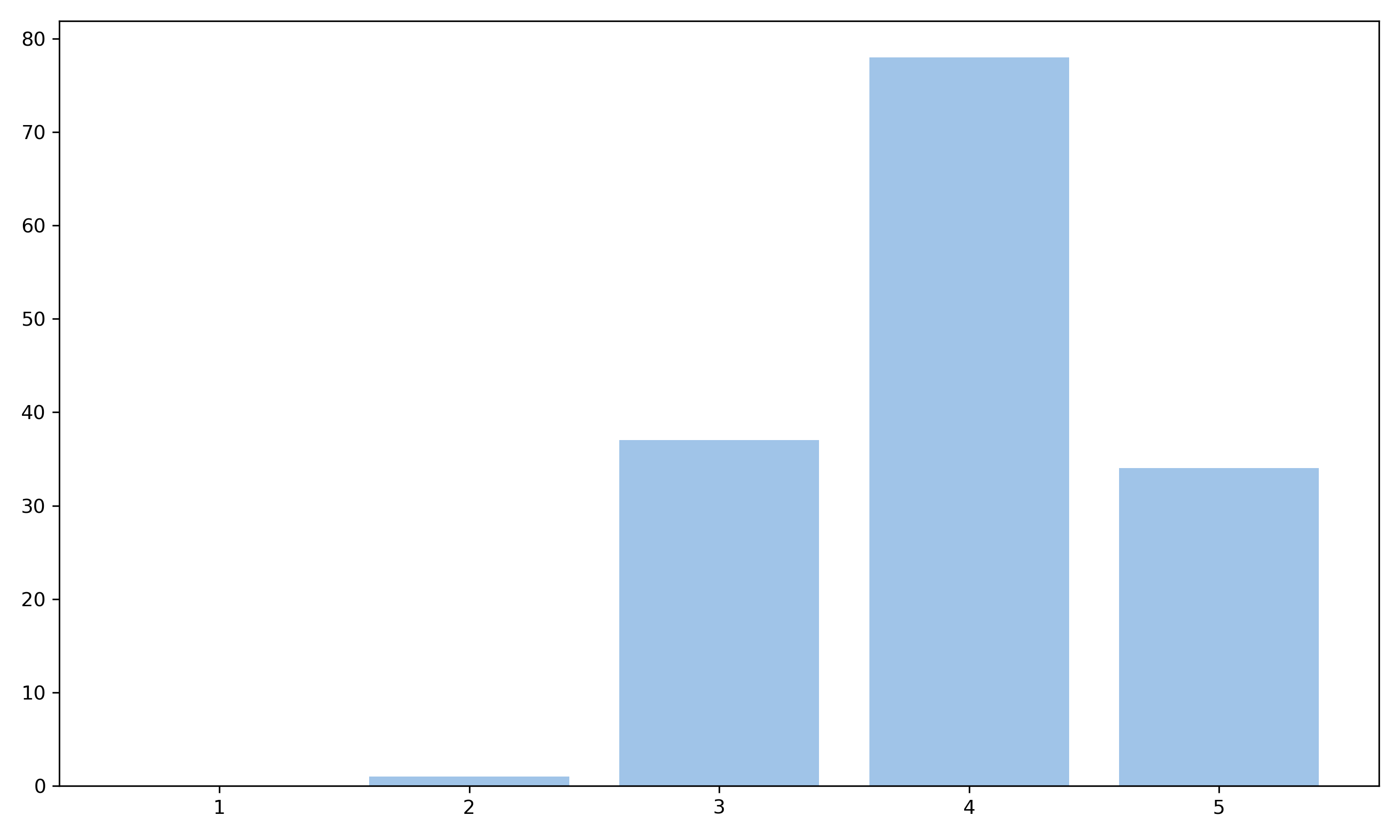}} \\
$\textit{S}_{13}$  &\makecell[l]{\textbf{[Record]} The date fields of version date in PriBOM would help in maintaining a record of privacy-related updates.} & 3.89 & \raisebox{-0.32\totalheight}{\includegraphics[width=0.04\textwidth]{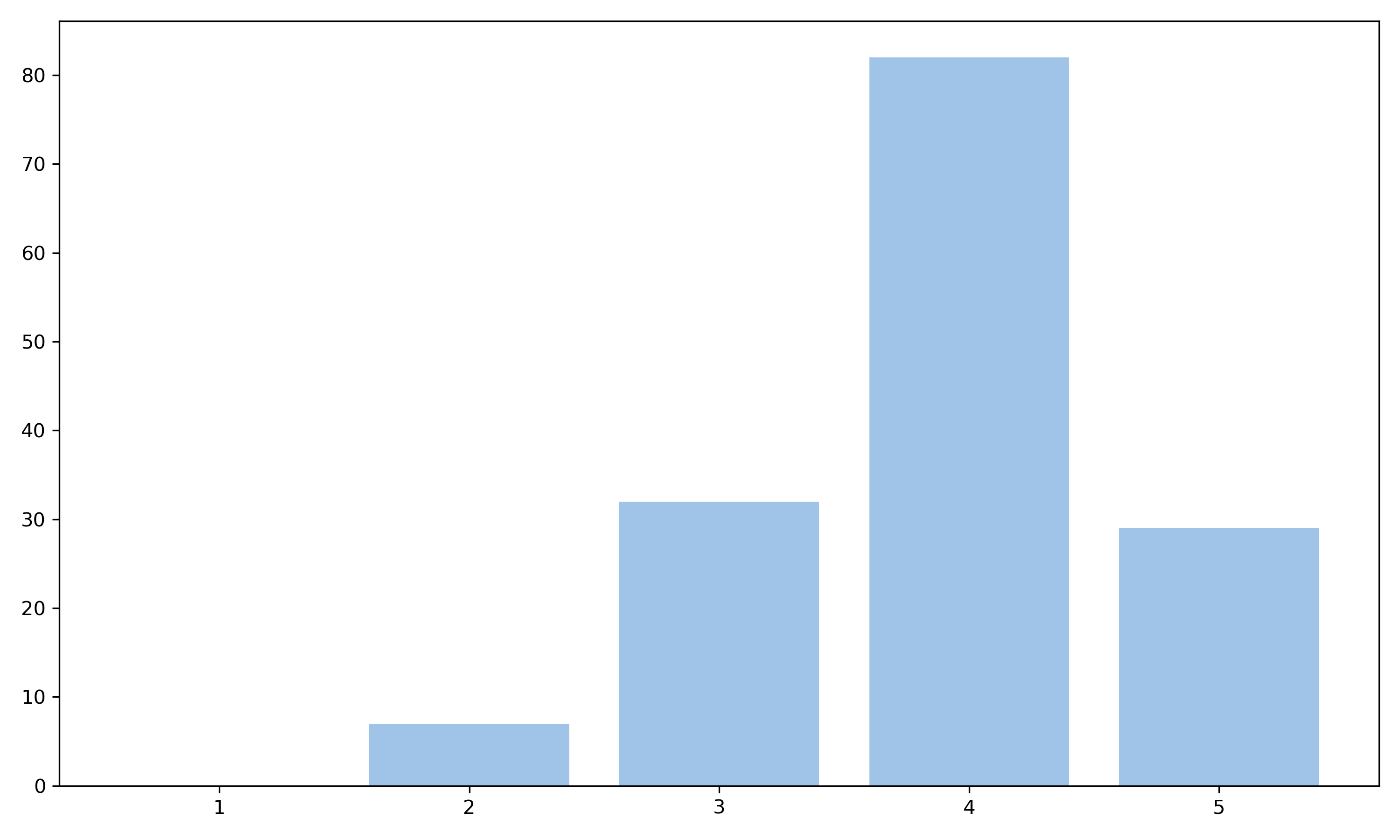}} \\
$\textit{S}_{14}$  &\makecell[l]{\textbf{[Update Awareness]} The data fields of version date in PriBOM would help developers aware of the TPL updates about privacy practices.} & 3.87 & \raisebox{-0.32\totalheight}{\includegraphics[width=0.04\textwidth]{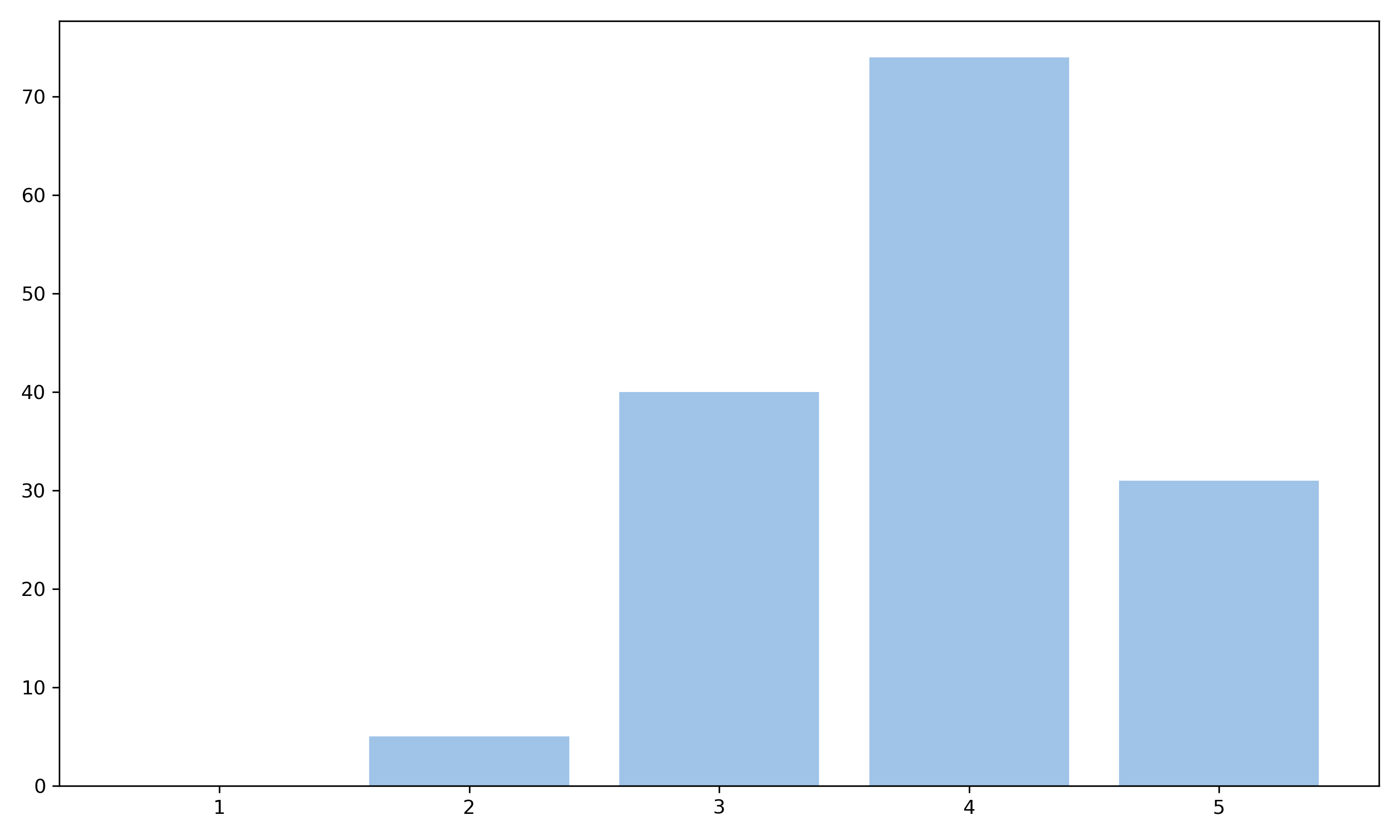}} \\
\midrule

\multicolumn{4}{c}{\textbf{Design of the Privacy Notice Disclosure Section of PriBOM}} \\

\midrule
$\textit{S}_{15}$  &\makecell[l]{\textbf{[Alignment]} The Privacy notice fields in PriBOM adequately guides the incremental development of accurate privacy notices.} & 3.71 & \raisebox{-0.32\totalheight}{\includegraphics[width=0.04\textwidth]{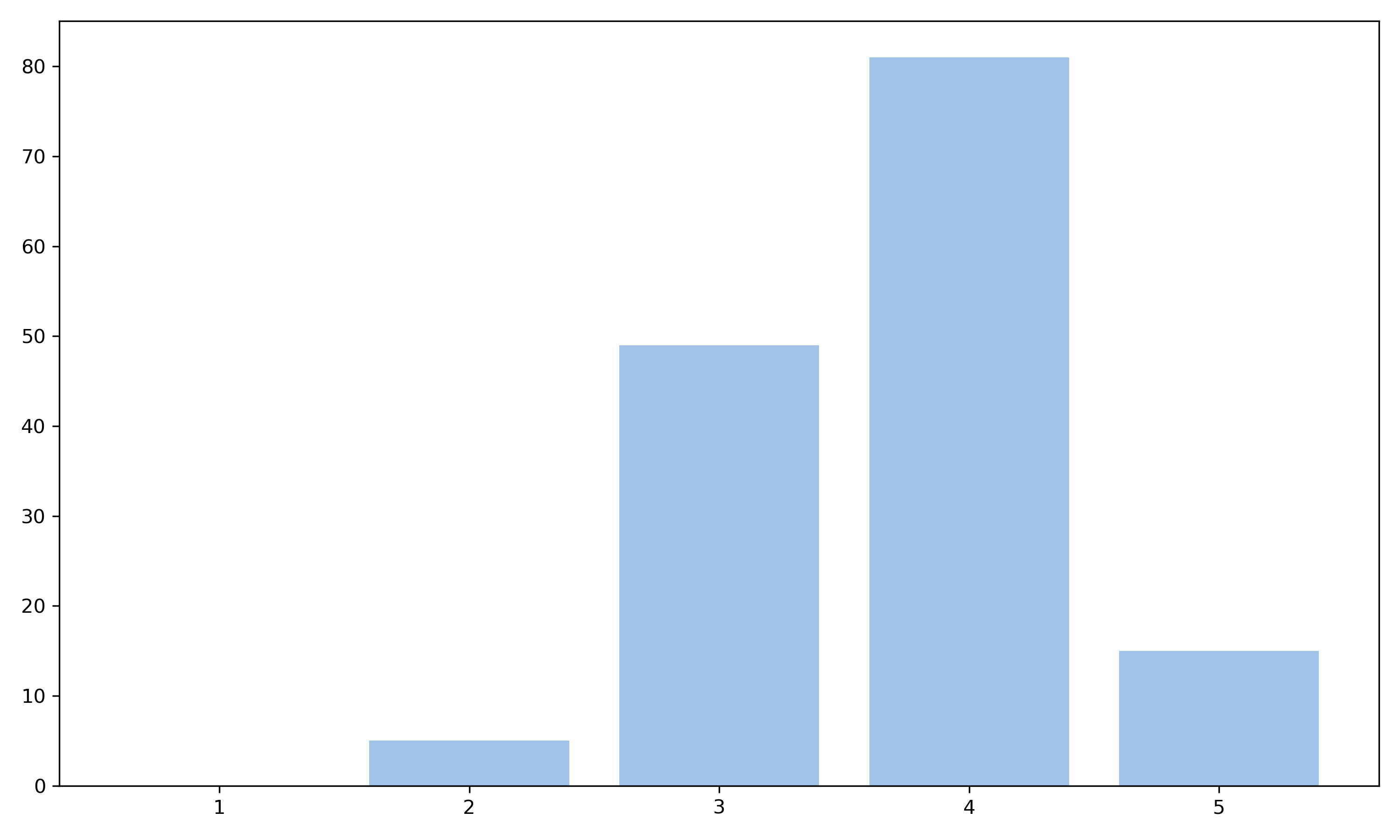}} \\

$\textit{S}_{16}$  &\makecell[l]{\textbf{[Traceability]} This section in PriBOM can help trace data practices to relevant descriptions in privacy policy.} & 3.85 & \raisebox{-0.32\totalheight}{\includegraphics[width=0.04\textwidth]{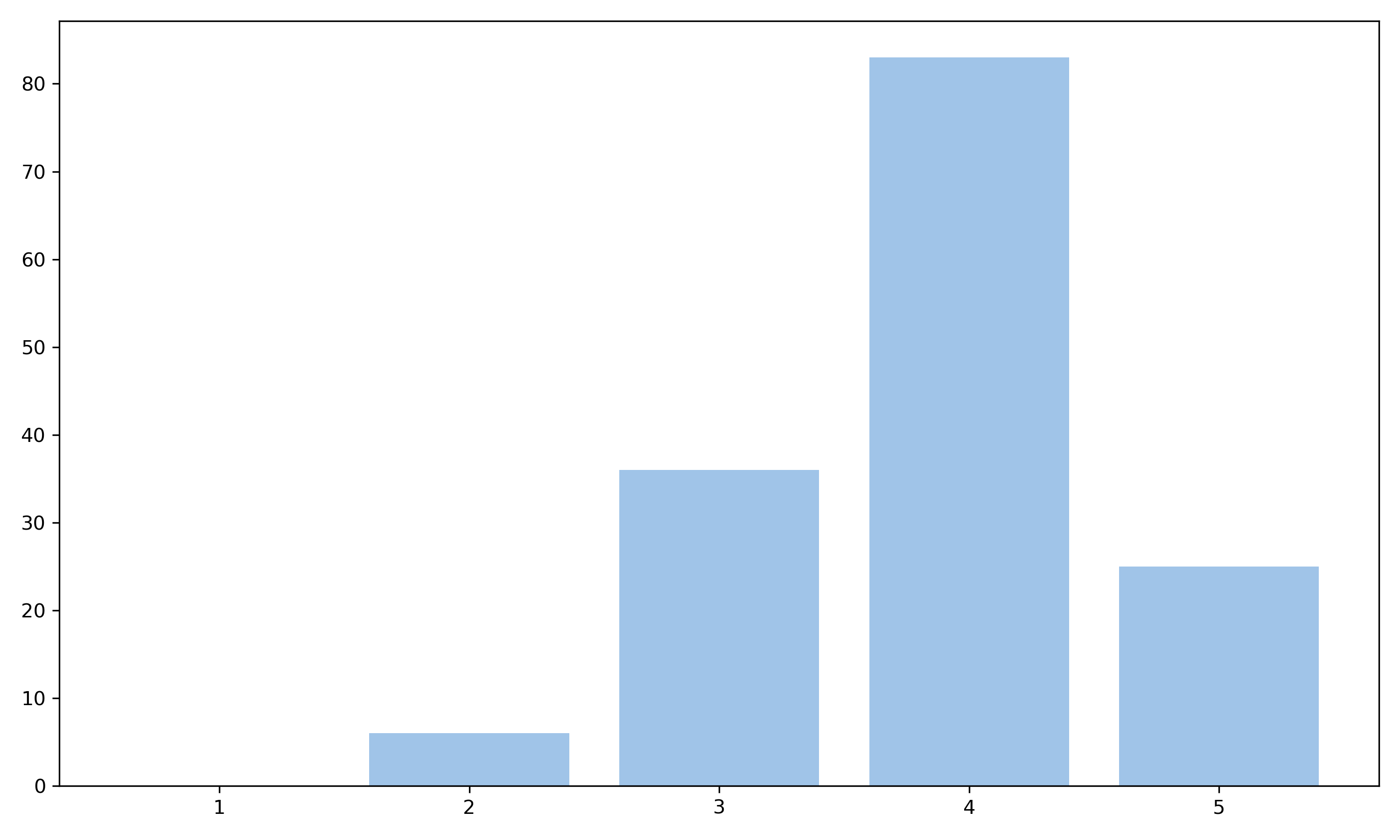}} \\
$\textit{S}_{17}$  &\makecell[l]{\textbf{[Trackability]} This section in PriBOM can facilitate tracking from disclosures in privacy policy to related data practices in code.} & 3.83 & \raisebox{-0.32\totalheight}{\includegraphics[width=0.04\textwidth]{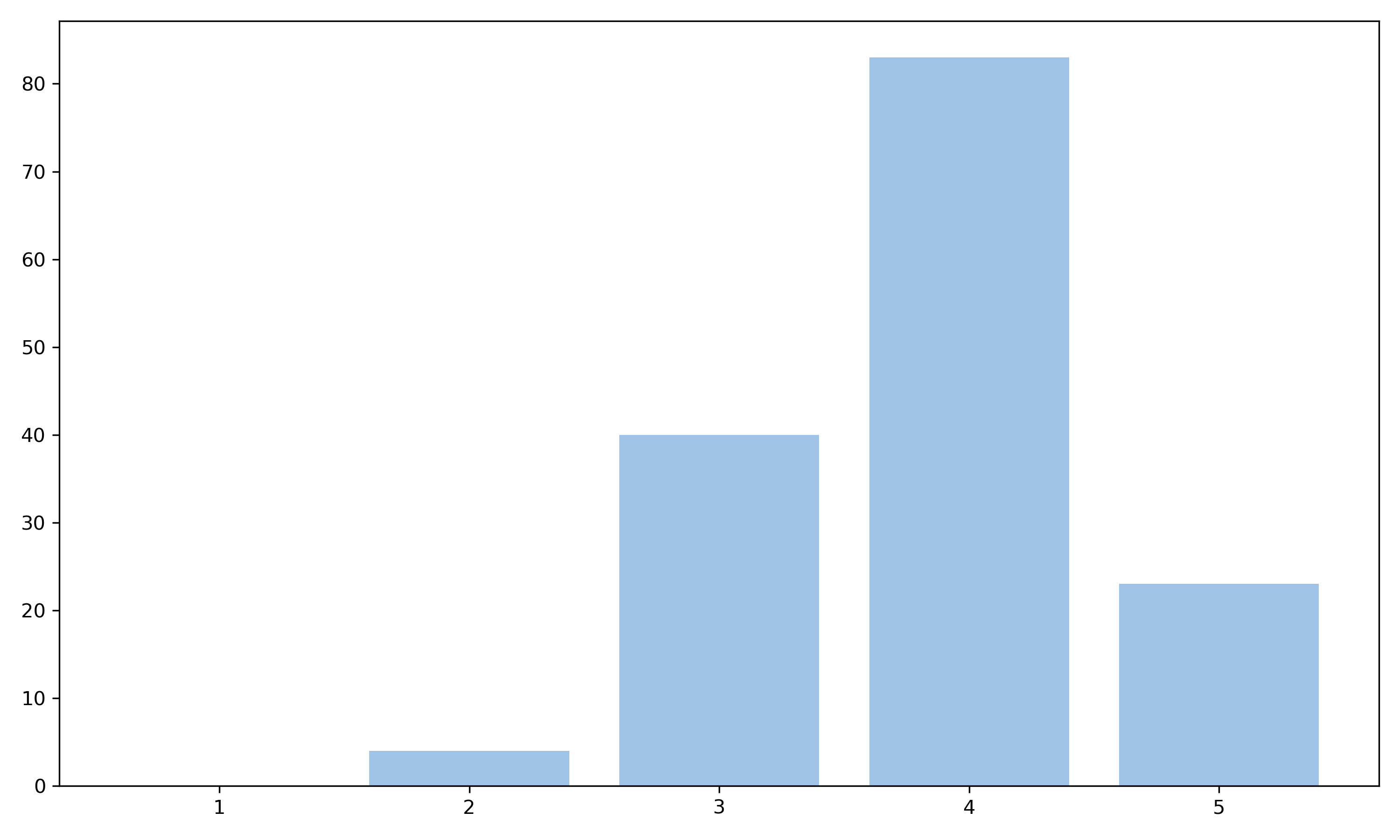}} \\
\midrule

\multicolumn{4}{c}{\textbf{Usability and Practicality of PriBOM}} \\

\midrule
$\textit{S}_{18}$  &\makecell[l]{\textbf{[Communication]} The PriBOM is a practical solution for efficient privacy-related communication between different roles in development team.} & 3.94 & \raisebox{-0.32\totalheight}{\includegraphics[width=0.04\textwidth]{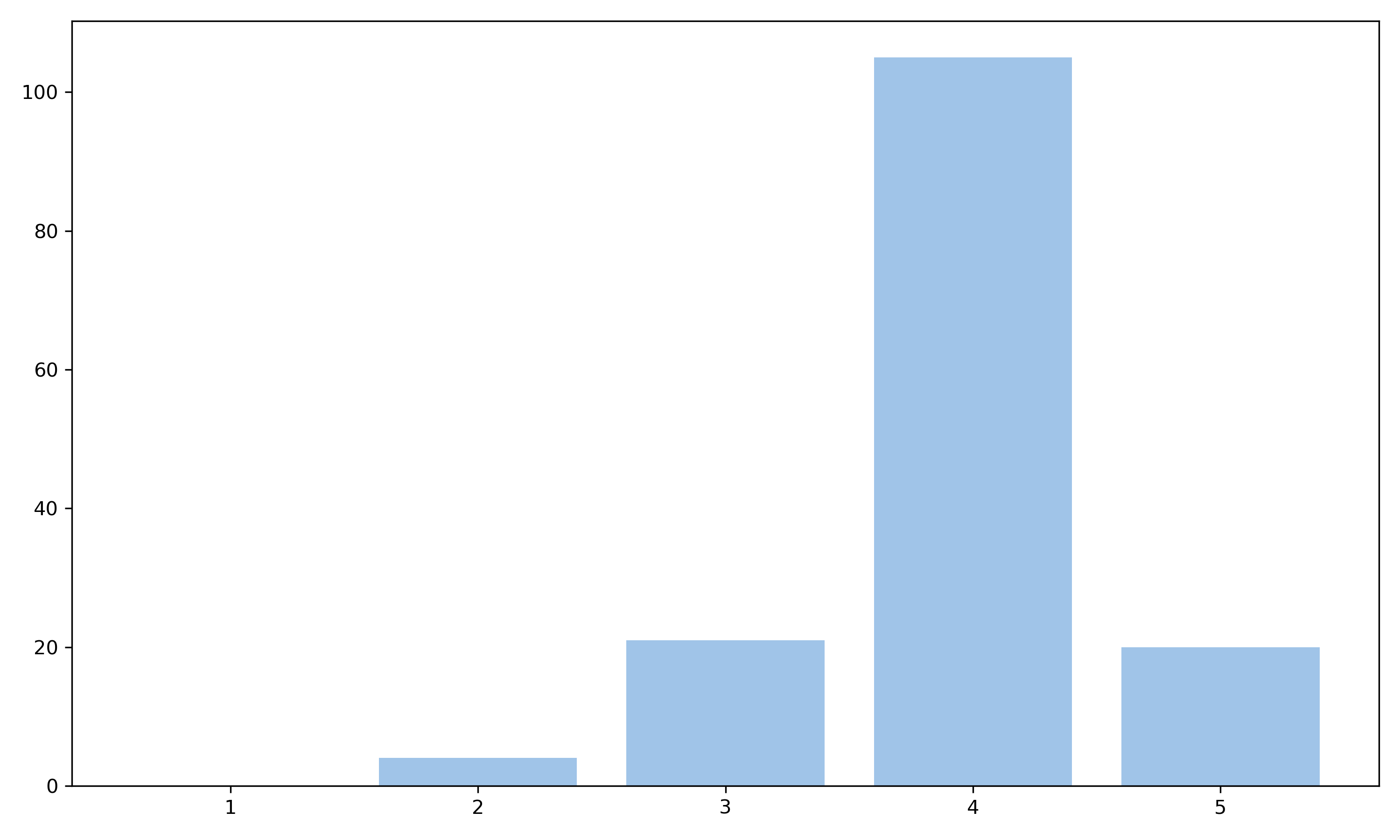}} \\
$\textit{S}_{19}$  &\makecell[l]{\textbf{[Privacy Notice Generation]} The PriBOM can streamline privacy notice generation and management for development teams.} & 3.92 & \raisebox{-0.32\totalheight}{\includegraphics[width=0.04\textwidth]{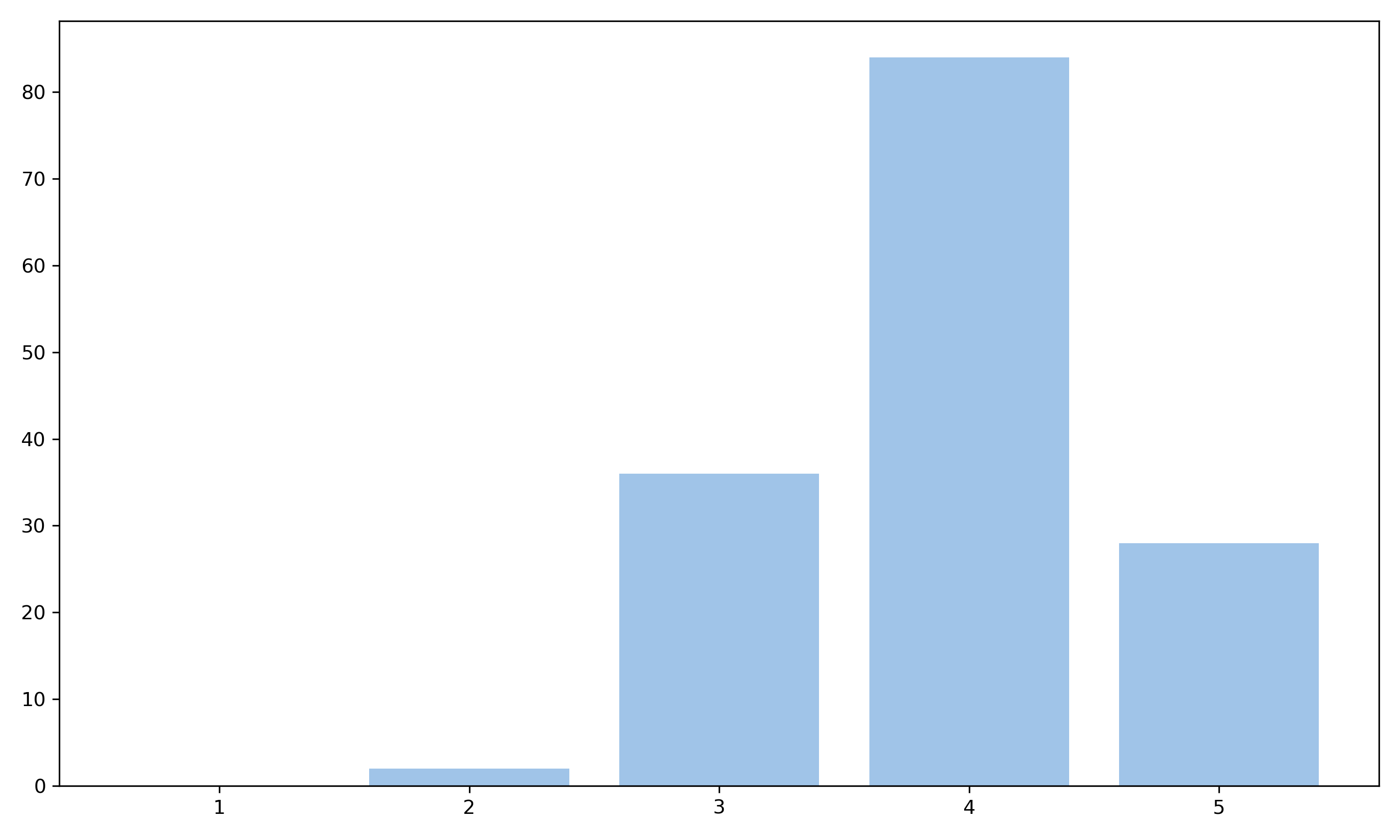}} \\
$\textit{S}_{20}$  &\makecell[l]{\textbf{[Disclosure-Behaviour Alignment]} The PriBOM would make it easier to align privacy notice pieces with app's actual software behaviours.} & 4.02 & \raisebox{-0.32\totalheight}{\includegraphics[width=0.04\textwidth]{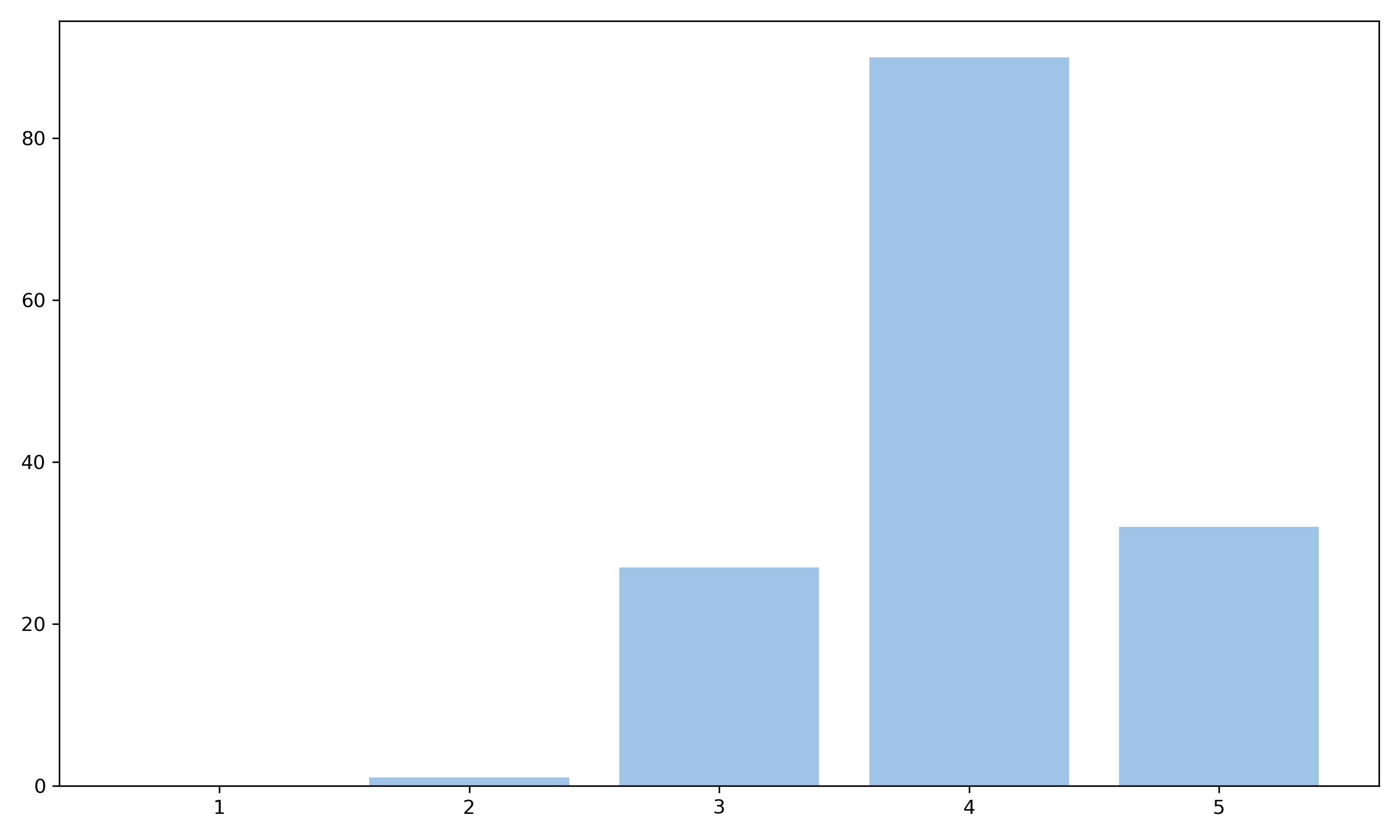}} \\
$\textit{S}_{21}$  &\makecell[l]{\textbf{[Transparency Promotion]} The PriBOM can enhance the transparency about data handling within development teams.} & 4.03 & \raisebox{-0.32\totalheight}{\includegraphics[width=0.04\textwidth]{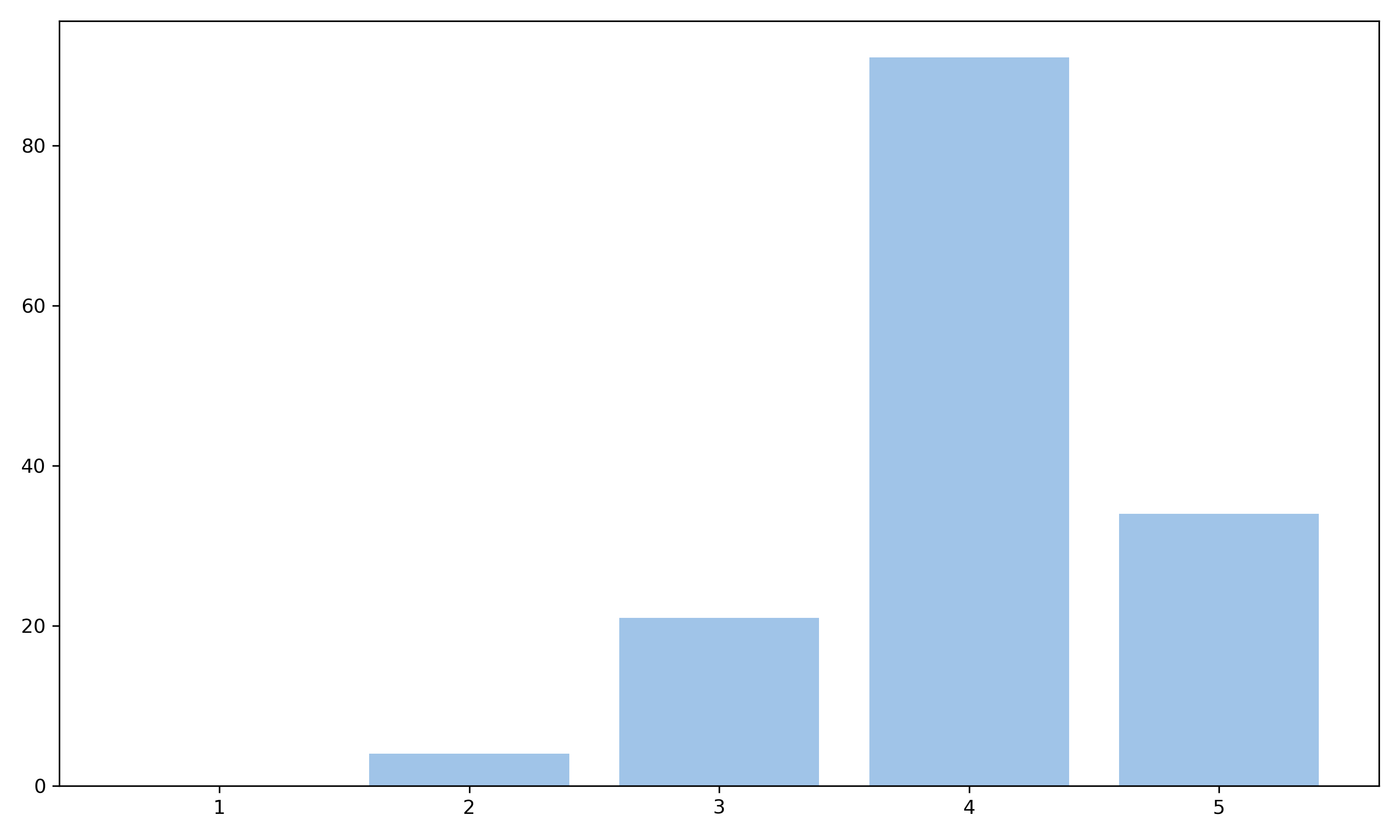}} \\
$\textit{S}_{22}$  &\makecell[l]{\textbf{[Privacy Awareness]} Implementing the PriBOM could lead to improved privacy awareness among the development team members.} & 3.99 & \raisebox{-0.32\totalheight}{\includegraphics[width=0.04\textwidth]{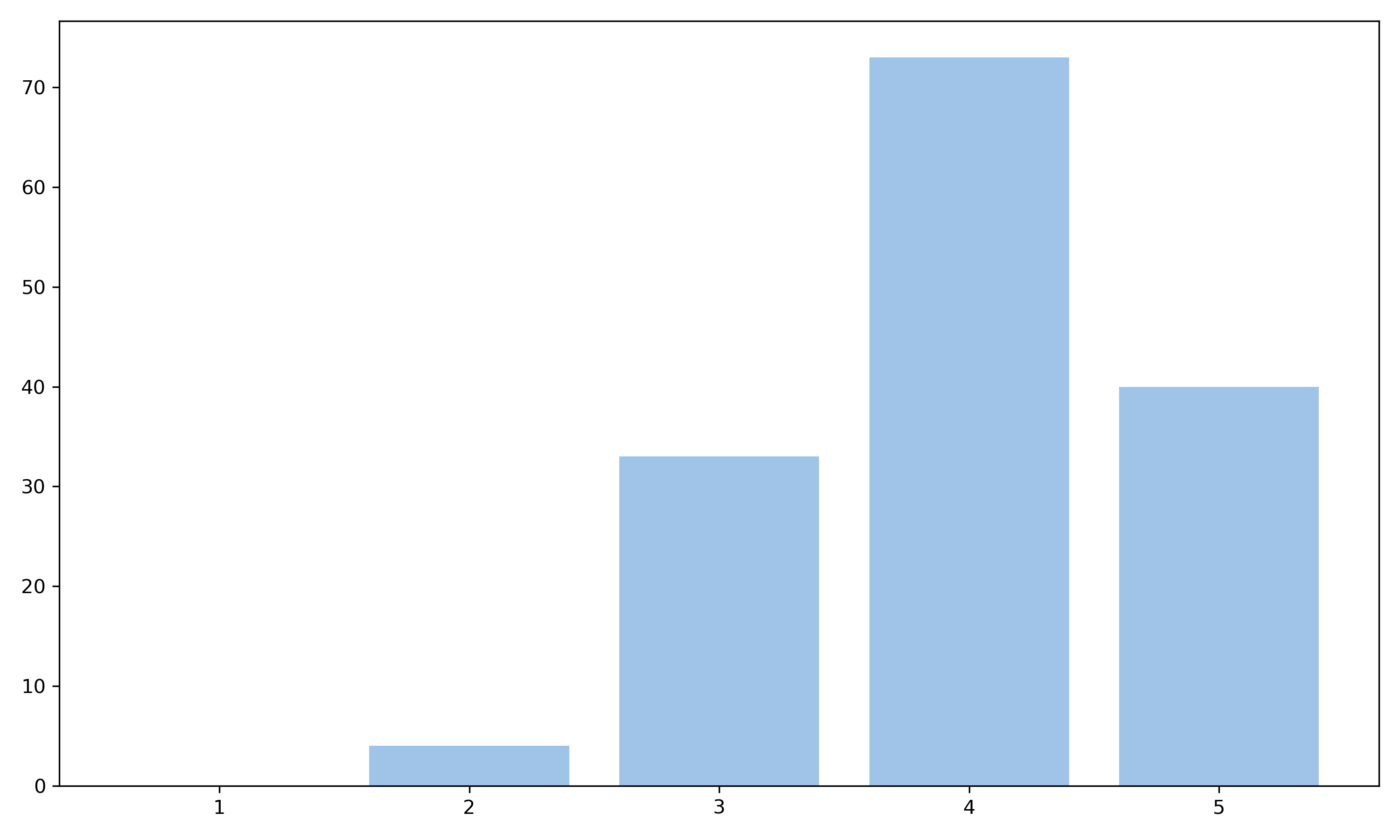}} \\
$\textit{S}_{23}$  &\makecell[l]{\textbf{[Risks Mitigation]} The PriBOM could systematically reduce the risks of overlooking privacy concerns during the development process.} & 3.96 & \raisebox{-0.32\totalheight}{\includegraphics[width=0.04\textwidth]{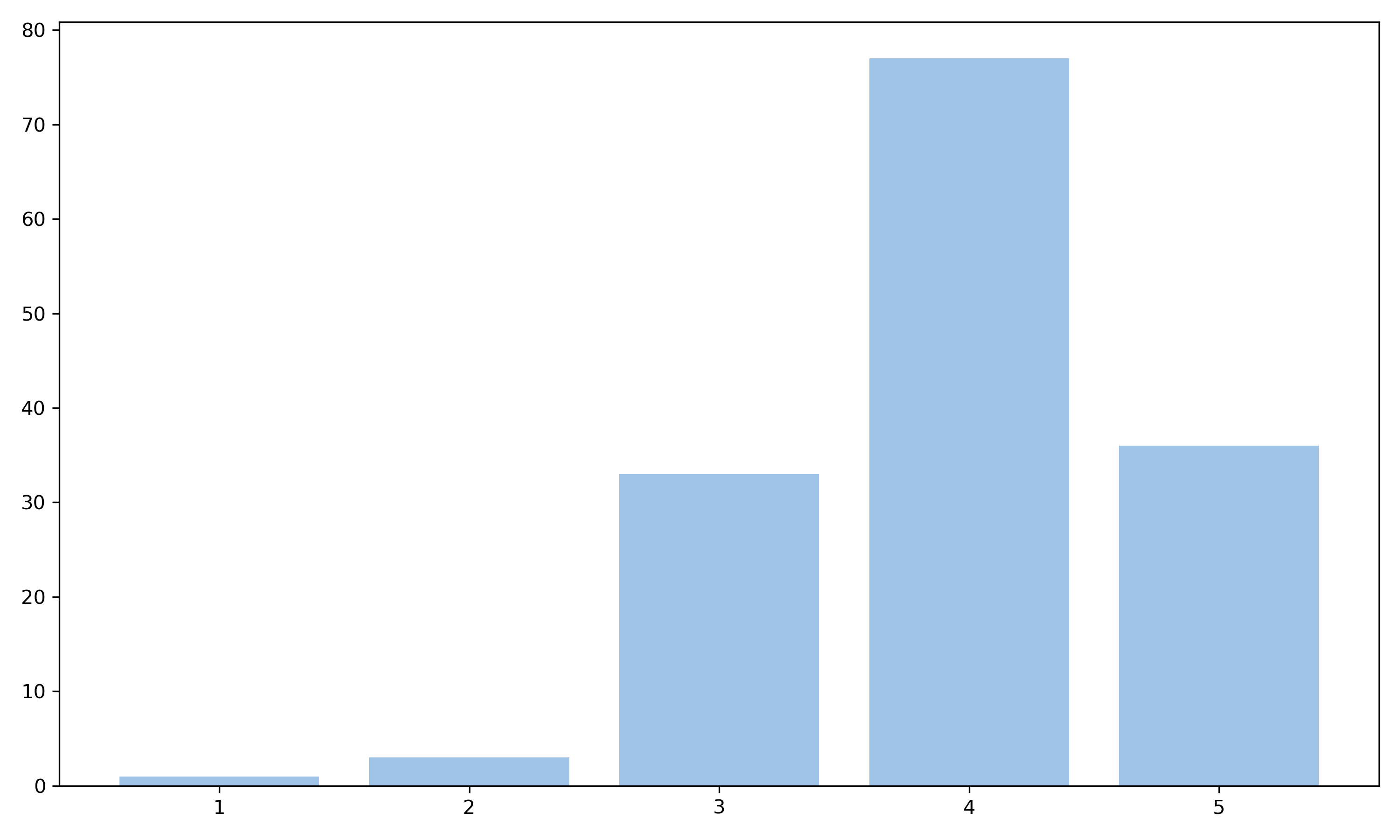}} \\
$\textit{S}_{24}$  &\makecell[l]{\textbf{[Scalability]} The PriBOM is a scalable solution that could be adapted for different project sizes and complexities.} & 3.69 & \raisebox{-0.32\totalheight}{\includegraphics[width=0.04\textwidth]{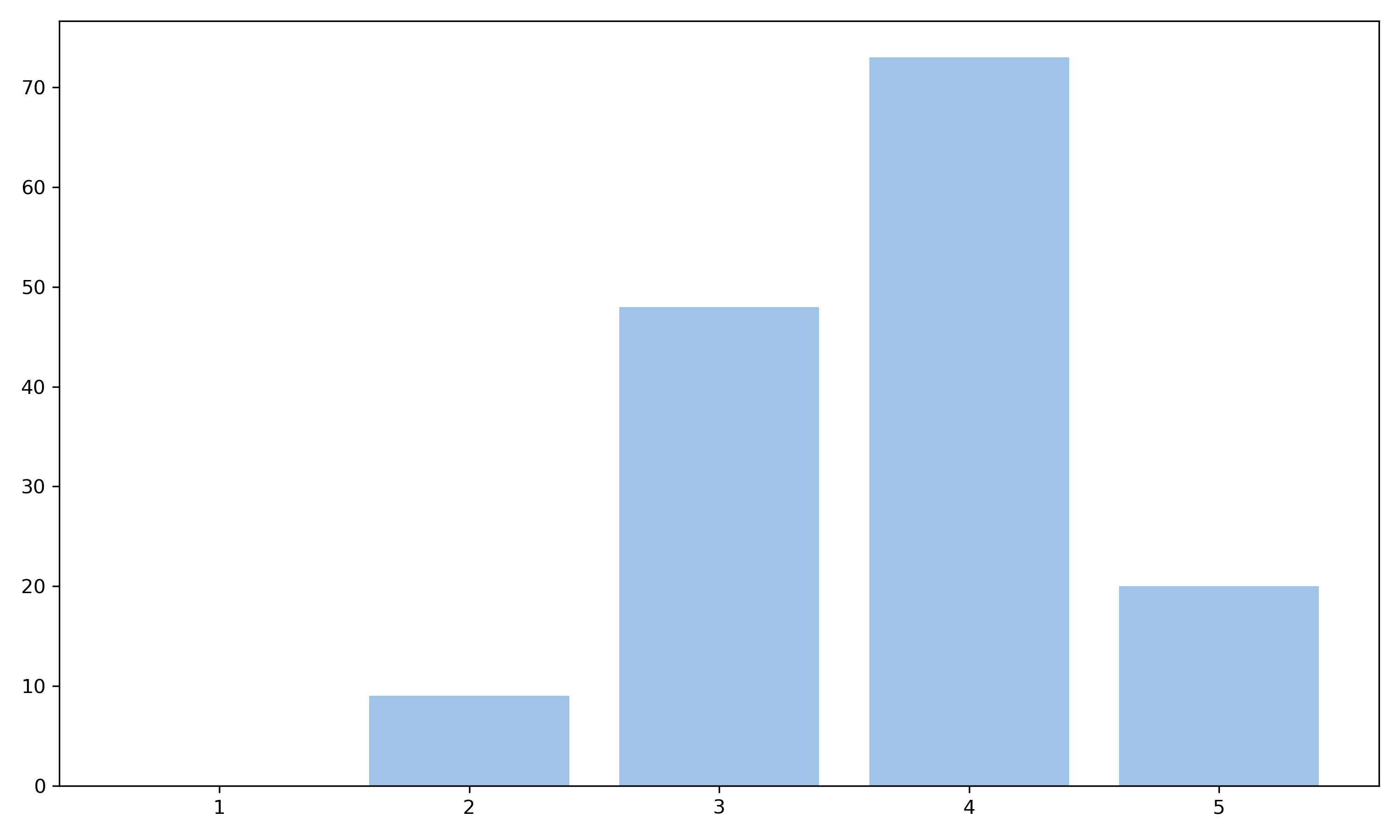}} \\
$\textit{S}_{25}$  &\makecell[l]{\textbf{[Consent Identification]} The PriBOM effectively aids in identifying user-consent-required data collection practices.} & 3.81 & \raisebox{-0.32\totalheight}{\includegraphics[width=0.04\textwidth]{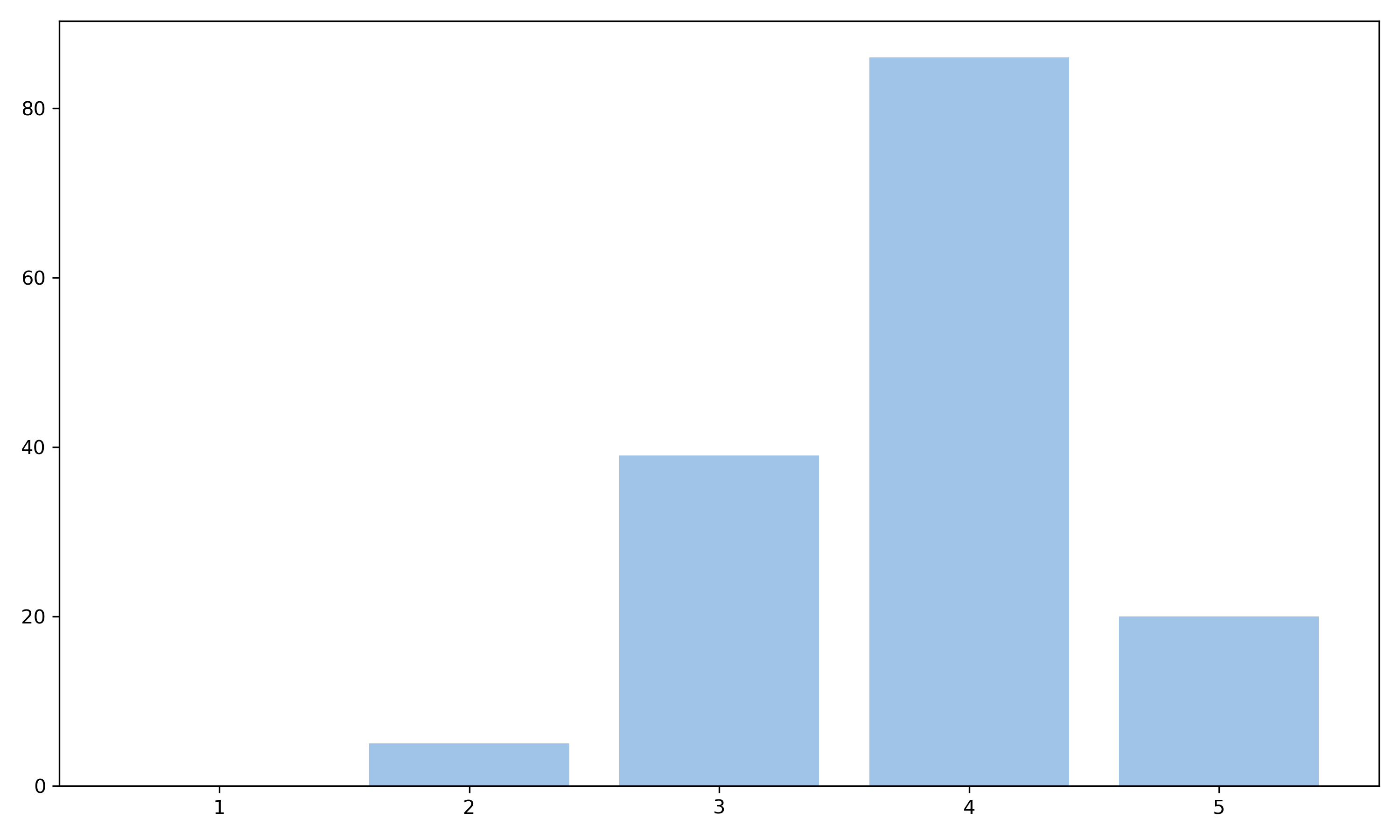}} \\
$\textit{S}_{26}$  &\makecell[l]{\textbf{[Inquiry Response]} PriBOM could reduce the effort needed to respond to privacy-related inquiries.} & 3.79 & \raisebox{-0.32\totalheight}{\includegraphics[width=0.04\textwidth]{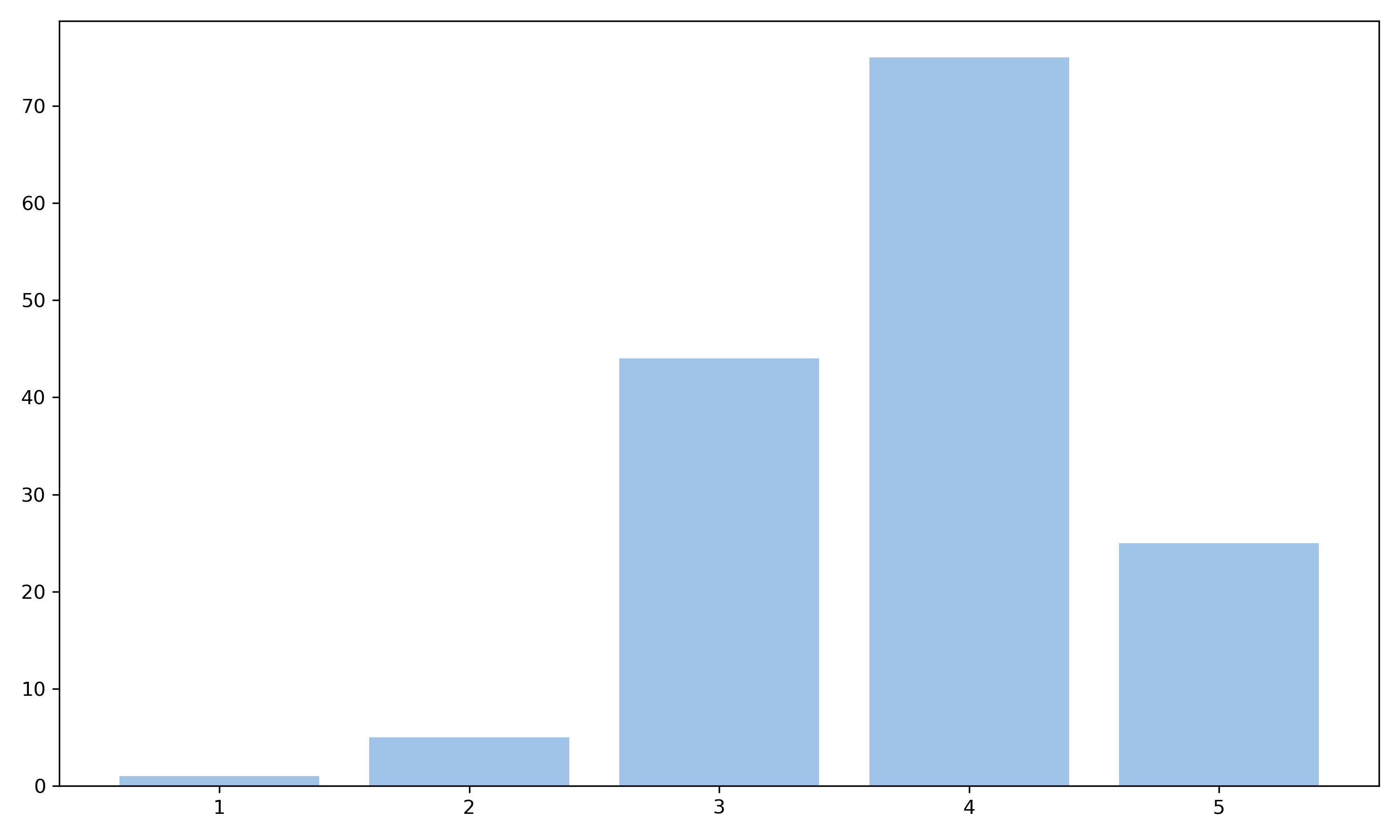}} \\

\bottomrule
\end{tabular}
}%
\end{table*}

%


\subsection{Participant Recruitment}
Following the participant recruitment processes in prior works in Usable Privacy and Security field~\cite{xia2023empirical, lin2023data, pan2023seeprivacy}, 
we published the survey via Qualtrics~\cite{Qualtrics}, and recruited participants through Prolific~\cite{Prolific}, a commonly used recruitment platform of sustainably gathering survey responses at scale. We use the participant screener in Prolific to recruit participants that meet our requirement discussed in Section~\ref{survey_design}.


We consider responses that are uncompleted or with a completion time of less than 2 minutes to be invalid. In total, we received 150 valid survey responses from participants with diverse backgrounds. 
The median time to complete the survey is 14 minutes and 33 seconds. 
Our participants span across 35 countries include 94 males and 55 females, with one participant preferring not to say, which conforms the reality that there are more male practitioners in software industry.
We also include 3 demographic questions in the survey questionnaire to understand the roles of participants in software development, years of experience, and the size of the team they work in. An overview of the demographic information is presented in Table~\ref{tab_demo}. The role with the highest proportion is junior developer (39\%), while other participants reported to be UI designers (15\%), project managers (14\%), senior developers (13\%), legal team members (9\%), and other roles such as CEO. 
The team size among participants is also diverse. The most common team size is less than 10 people (59\%), while there are also 9\% of participants working in a team with more than 50 people. We believe our participants' backgrounds are sufficiently diverse for our survey. 
For ethics consideration, please refer to Section~\ref{Ethics}.




\subsection{Ethics Considerations}
\label{Ethics}
We recruited participants through Prolific~\cite{Prolific}, a highly regarded online platform for researchers to recruit participants for UPS studies.
We only collect basic and non-identifiable demographic information.
Participants receive monetary rewards at a rate of £7.49 per hour, as recommended by Prolific.
We ensure the risk of participating in this research is minimum, and no disturbing content is distributed in the survey.
A consent and withdrawal information sheet was presented before the survey to inform participants of their rights, including the ability to withdraw at any point before submitting by exiting the survey page.
Participants give their consent by proceeding with the survey. 
All information provided by participants is treated confidentially and we will not share the collected data to any third parties.

%
\begin{table*}[t]
\centering
\caption{The free text open-ended questions in our survey.}
\vspace{-5pt}
\label{tab_open_question}
\resizebox{1.0\textwidth}{!}{%
\begin{tabular}{l|l}
\toprule

\textbf{No.} & \textbf{Open-Ended Question} \\

\midrule

$\textit{Q}_{1}$ & \makecell[l]{How would implementing PriBOM in your projects impact the management and documentation of permission request and data collection? Please briefly describe\\ any potential advantages or challenges.} \\
\midrule
$\textit{Q}_{2}$ & \makecell[l]{Reflecting on your previous projects, how would detailed tracking of Third-Party Libraries (TPLs) as proposed by PriBOM facilitate to manage privacy compliance?} \\
\midrule
$\textit{Q}_{3}$ & \makecell[l]{What potential benefits do you see in the PriBOM method of documenting data practices and their associated disclosures in the privacy notice?} \\
\midrule
$\textit{Q}_{4}$ & \makecell[l]{Reflecting on your own experiences, how do you think implementing PriBOM in the development team might influence the collaboration between different roles\\ toward privacy notice (e.g., privacy policy) management and generation?} \\

\bottomrule
\end{tabular}
}%
\end{table*}
%

\subsection{Result Analysis}
The results of the survey are detailed in Table~\ref{tab_statements}. We calculate the average agreement score and present the distribution of Likert-scale responses. 
Overall, participants largely agree with our statements regarding \texttt{PriBOM}. 
For open-ended responses, we employ the methodology of deductive thematic analysis~\cite{braun2006using, crabtree1992template} to identify key insights. Given that the free-text questions are already structured around specific sections of our survey (e.g., $\textit{Q}_{1}$ corresponds to Design of the Codebase and Permission Section in Table~\ref{tab_statements}), we align our qualitative analysis with predefined two-layer (sections and statements) a priori rather than emerging new codes and themes from responses.
Below, we discuss the results in more detail.

\subsubsection{Design of \texttt{PriBOM}} 
\label{design_of_Pri}
Most participants agree or strongly agree with the statements related to the general design. The intuitiveness of \texttt{PriBOM}'s design is well received with 85.33\% agreement, suggesting that most participants found the data fields in \texttt{PriBOM} logically organized and conducive to understanding. The responses regarding the precision and clarity of Widget Identification show a trend toward agreement. Notably, most participants (83.33\%) agree or strongly agree \texttt{PriBOM} is helpful in providing clear terminology. 

The results of the Codebase and Permission Section suggest a favorable perception among participants, particularly in their recognition of \texttt{PriBOM}'s functional contributions to privacy management. The positive attitude (80.67\%) toward the inclusion of \textit{Permission} and \textit{Data Type} fields highlight their importance. Interestingly, the statement regarding the necessity of permission information in improving data transparency received the highest average agreement score (4.09) among all statements. \texttt{PriBOM}'s approach has also been recognized. For example, one participant stated ``\textit{I believe a system like this would be helpful to give our clients more peace of mind on the quality of our privacy policies}'' ($\textit{Q}_{1}$), indicating participants believed \texttt{PriBOM} was useful for creating accurate privacy policies. 

The results of the TPL Section illustrate a strong positive perception regarding \texttt{PriBOM}'s capabilities in managing TPL information. 74.67\% of the participants agree or strongly agree that documenting TPL
versions may help identify discrepancies in privacy practices between different TPL versions. One participant wrote ``\textit{I believe the key terms here are the dates and update version records. With privacy laws changing all the time, it would be a way developers could track the changes necessary to TPLs to ensure compliance}'' ($\textit{Q}_{2}$). Participants also expressed their agreement with the trackability of \texttt{PriBOM}, as stated, ``\textit{...if this is done properly, you have a \underline{concrete evidence} where things went wrong}''. Such support could relieve the challenge of limited technical knowledge discussed in Section~\ref{sec_formative_study}. 

The results of the Privacy Notice Disclosure Section also indicate a positive perception, particularly in terms of traceability (72\%) and trackability (70.67\%). As participants mentioned in $\textit{Q}_{3}$, ``\textit{I believe the tracking and traceability components benefit the user by providing \underline{clear and concise management opportunities} for accuracy}''. Moreover, the role of \texttt{PriBOM} in regulatory compliance of applications has been supported, as ``\textit{It ensures that the developers access and/or control the information \underline{in line with the law or regulations set}}''. Since the privacy notice generation involves collaboration, the collaborative nature of \texttt{PriBOM} has also been recognized, ``\textit{Seems like the greatest benefit of this, in particular, is to facilitate communication between devs and other departments}''. Such collaboration with legal teams helps address the challenges of privacy knowledge absence.

As for the usability and practicality, the results indicate strong approval (83.33\%) for \texttt{PriBOM} in enhancing privacy-related communication across different development roles. Stated in $\textit{Q}_{4}$, ``\textit{I think it will \underline{influence a culture} within the workspace that is focused on privacy and security and creating awareness in this regard}'', indicating that \texttt{PriBOM} may address the organizational environment challenges discussed in Section~\ref{sec_formative_study}. The role of \texttt{PriBOM} toward creating privacy policy is also well recognized:
\vspace{3pt}
\quoteFrame{``\textit{this common language investors clear communication and understanding among team members, facilitating more efficient collaboration in drafting and updating privacy policies}.''}

\begin{tcolorbox}[boxsep=1pt,left=2pt,right=2pt,top=2pt,bottom=2pt, boxrule=0mm]
\textbf{Highlight-1:} Participants highly value the intuitiveness and clarity of \texttt{PriBOM}'s design, with high agreement on its capability to provide structured data fields and transparent privacy information.
Results affirm that \texttt{PriBOM} significantly bolsters privacy-related communication, exhibiting a high level of usability.
\end{tcolorbox}

\subsubsection{Viewpoint Comparison Across Roles}
\label{comp_diff_role}
During the data analysis, we observed that different roles hold different views on \texttt{PriBOM}. To further investigate this difference, we separated the participants based on their roles and calculated the agreement scores for each group on the statements separately. Although different roles showed similar agreement in many statements, we concluded two findings.

\begin{figure}[t]
  \centering
  \includegraphics[width=.98\linewidth]{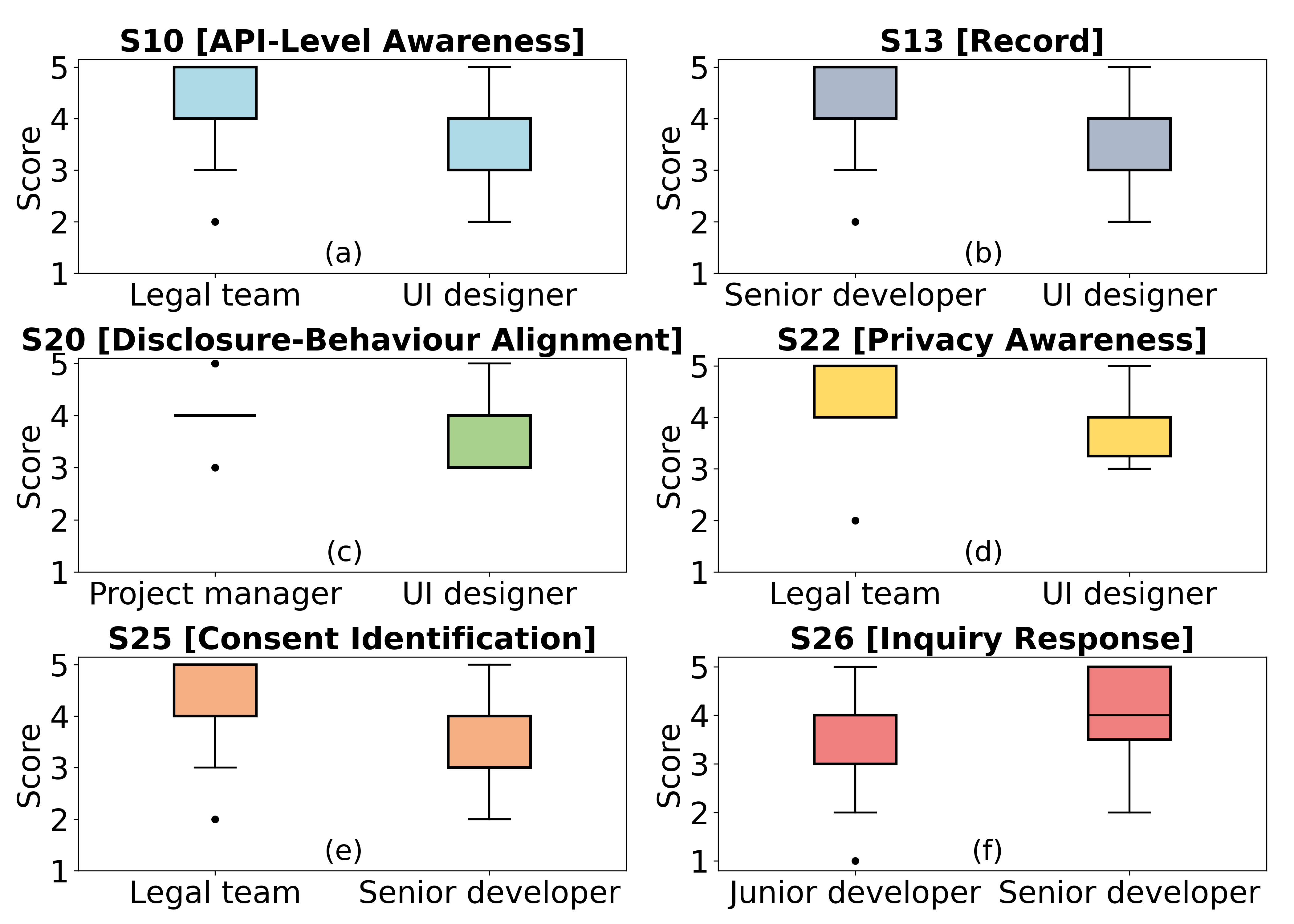}
    \caption{Examples of the differences of agreement scores regarding survey statements among different roles.}
  \label{fig_role_boxplot}
\end{figure}

The first finding is the discrepancies in the perspectives between frontend UI designers and other roles. 
For instance, The legal team's agreement score for $\textit{S}_{10}$ is significantly higher than that of UI designers, as shown in Figure~\ref{fig_role_boxplot}(a). This statement describes the necessity of API-level information for privacy information management. The legal team's average score reached 4.00, while the average score for UI designers was 3.63, indicating a divergence of opinions between them, with the former largely agreeing on it and the latter not. We believe this result is reasonable since legal experts typically have a higher privacy awareness, while UI designers may consider this issue less. This relatively low level of agreement may lead to UI designers lacking a rigorous and serious attitude towards privacy when dealing with such information.

Similarly, UI designers scored an average of 3.82, while legal teams scored significantly higher at 4.31 on the statement concerning the impact of \texttt{PriBOM} on enhancing privacy awareness among team members (Figure~\ref{fig_role_boxplot}(d)). 
This means that though legal experts may see \texttt{PriBOM} as a critical tool for privacy practicing, UI designers might appreciate its role but to a lesser extent, as their direct interaction with privacy as a compliance or educational tool is less pronounced. 
Distinct differences in perception between the project managers and UI Designers were observed in the disclosure-behaviour alignment aspect of \texttt{PriBOM} (Figure~\ref{fig_role_boxplot}(c)). The project managers, who need to ensure the development aligns with business and compliance objectives, show a higher agreement score of 4.19, suggesting a strong recognition of the utility of \texttt{PriBOM} in streamlining and clarifying the alignment between privacy disclosures and the actual behaviors.
Regarding the maintenance of a record of privacy-related updates (Figure~\ref{fig_role_boxplot}(b)), UI designers gave a lower score of 3.64, compared to senior developers who scored it at 4.16.
Most discussions of privacy on Reddit were triggered by external events ~\cite{li2021developers}, including updates of third-party code that is used. While senior developers are likely to value features that aid in documenting and archiving changes, which supports better version control and compliance tracking, UI designers may not perceive it as impactful to their role.

While UI designers often give scores that differ significantly from other raters, differences in opinion also occur between other roles. For instance, in the case of aiding in identifying practices that require user consent (Figure~\ref{fig_role_boxplot}(e)), legal teams rated \texttt{PriBOM}'s effectiveness at 4.00, whereas senior developers gave it a lower score of 3.68. This difference can be attributed to the legal team's acute need for precise tools in drafting and verifying privacy policies and consent protocols. Conversely, one of senior developers, who often focus on actual coding following privacy rules, commented ``\textit{I think it is complicated to follow these steps...}''. 

Our second finding is differences in experience can lead to changes in viewpoints. For example, senior developers agree more than junior developers that \texttt{PriBOM} could reduce the effort needed to respond to privacy-related inquiries ($\textit{S}_{26}$), as presented in Figure~\ref{fig_role_boxplot}(f). This statement relates to whether \texttt{PriBOM} could provide support in addressing privacy inquiries. Since \texttt{PriBOM} serves as a privacy information inventory, we believe the development team can respond faster based on the information in PriBOM, rather than manually extracting them after receiving requests. The senior developers' average score was 4.05, while the average score for UI designers was 3.78, suggesting that senior developers largely agree with our statement, while junior developers may not. We believe this could be due to the more experience of senior developers who may have a better understanding of privacy inquiry and the role \texttt{PriBOM} can play in it. Differently, junior developers may not have much exposure to privacy inquiries, and the uncertainty may guide them to choose more neutral options.


Overall, the human evaluation of \texttt{PriBOM} across multiple dimensions has garnered largely positive feedback, underscoring its usefulness. However, participants also provided feedback on the potential challenges of \texttt{PriBOM} in practical adoption. 
\textit{``It would be challenging to let my company's elderly employees get up to date on this since most of them are \underline{stuck to their old ways}}.'' Another participant also wrote ``\textit{it would be a challenge and a matter of time for everyone to learn how to work based on PriBOM}''. These feedbacks indicate that \texttt{PriBOM} should continue to improve usability and reduce the learning curve. Nevertheless, participants still expressed affirmation of PriBOM, demonstrating that \texttt{PriBOM} is well-received by the development community for its comprehensive approach to embedding privacy into the software development lifecycle, as 
\vspace{3pt}
\quoteFrame{``\textit{I think it's a good Model that deserves to be tested, it may have a very positive impact and improve the way we look at privacy policies and management of projects in future}.''}

\begin{tcolorbox}[boxsep=1pt,left=2pt,right=2pt,top=2pt,bottom=2pt, boxrule=0mm]
\textbf{Highlight-2:} Although \texttt{PriBOM} is widely recognized for its advantages, perceptions of its value differ notably across various roles. Experience within the same role can further shape these viewpoints.
\end{tcolorbox}

\subsubsection{Perspectives Held by Non-technical Roles}
\label{view_non_tech}

\texttt{PriBOM} provides a systematic solution for communication and collaboration between technical and non-technical roles. 
As we discussed in the formative study, existing research has intensively discussed the opinions held by technical roles.
To understand the perspectives of non-technical roles, namely the legal team, UI designers, and project managers, on \texttt{PriBOM}, we analyzed their free-text responses and summarized the frequently mentioned topics. 

\textbf{Efficiency.} All three non-technical roles mentioned the efficiency benefits of \texttt{PriBOM}. What the legal team cares about is that \texttt{PriBOM} can reduce their heavy workload and speed up their work. A legal expert working in a team of 10 to 20 people commented ``\textit{Our team can get a quick guide to be complient with all the ness...PriBOM will help us with fast answer...}'', and another legal expert also stated ``\textit{it could speed up and automatize some processes}''. Differently, two UI designers mentioned ``\textit{high rates of turnover within projects}'' and ``\textit{team dynamics}'', which means team members may frequently join or leave the project due to various reasons such as changing job roles or shifts in project focus. In such environment, \texttt{PriBOM} could maintain consistency and enhance knowledge transfer by ensuring new members can quickly come up to speed on the existing privacy practices and policies without a steep learning curve. The project manager's thinking focuses on the role of \texttt{PriBOM} in decision-making. ``\textit{I like the clarity and effectiveness of it...}'', commented by a product manager working in a team with more than 50 people, 
\vspace{3pt}
\quoteFrame{``\textit{it would certainly ease decision making since it has very in depth record and makes it easier to trace things}''. }
All three non-technical roles mentioned challenges in the practical usage and adaptation of \texttt{PriBOM}. A legal expert in a team with over 50 people commented that ``\textit{Each department had its own unique ideas of how privacy notice works}'', and challenges pointing by UI designers include ``\textit{initial setup complexity, potential integration issues with existing systems}'' and ``\textit{it was something people would need to get used to first}''. Differently, the project manager's concern includes the effort required to maintain \texttt{PriBOM} within a large team. ``seems like a lot of effort to get started and to maintain it.'' Fulfilling \texttt{PriBOM} is indeed a chore, therefore we introduce a pre-fill that can save developers from adding extra burden.


\textbf{Traceability and Trackability.} 
This discussion mainly focuses on the benefits of constructing privacy chains for compliance analysis. As stated by a legal team member, ``\textit{It enhances the traceability in the privacy chain connecting front-ed UI and the product’s policy}''. One UI designer also mentioned ``\textit{efficient chain flows and better root cause analysis}'', and another one stated that ``\textit{It's like \underline{a trust-builder} because it ensures that what's promised in the notice \underline{matches up} with what's actually happening with their data, making compliance efforts smoother too}''. A product manager considered the communication between technical and non-technical roles involved in the privacy chain, and wrote 
\vspace{3pt}
\quoteFrame{``\textit{Imagine a central platform where developers can directly link data practices to specific components. This transparency would streamline communication between developers and privacy specialists}''.}

\textbf{Communication.} The discussion of this topic by the three roles focuses on different perspectives. Product managers think of responsibility and time management as key aspects. One product manager mentioned the benefit of ``\textit{clear and consistent segmentation of responsibility that impact liability}'', indicating his belief on \texttt{PriBOM} in facilitating responsibility assignment within the team. Another manager wrote ``\textit{improve time management, brings members to closer because they can see everything ...}'', supporting the use of \texttt{PriBOM} to improve communication efficiency.
For UI designers, they agree with the potential of \texttt{PriBOM} as a common communication platform, as ``\textit{It's like a \underline{common language} that everyone can understand, making it easier for developers, privacy officers, and legal teams to work together effectively}''. One participant from the legal team also mentioned ``\textit{I think we could get a better collaboration between the departments where we all are more clear about our roles}''. Overall, many responses support that the idea of \texttt{PriBOM}, a systematic solution, can help facilitate communication between different roles.

\textbf{Influence.} Several responses mentioned the influence of \texttt{PriBOM} on the software development environment. For example, one legal expert wrote ``\textit{It allows you to have a complete record which encourages healthy and smart data practices}'', indicating that \texttt{PriBOM} could promote a proactive approach to privacy management, encouraging teams to consider privacy implications continuously rather than as a periodic compliance exercise. One UI designer stated 
\vspace{3pt}
\quoteFrame{``\textit{I currently find the knowledge regarding privacy policy, permissions etc. to be lacking within the team, especially on a management level. I feel like it is currently regarded as an afterthought ... this could hopefully increase awareness among all team members and facilitate communication as well as implementing a proper workflow for managing privacy settings}''.}
\vspace{3pt}
This comment reveals a critical gap in privacy awareness and management at the management level within teams, where privacy considerations are not fully integrated into the development lifecycle. The need for a systematical solution like \texttt{PriBOM} for various roles is emphasized in such context.

\begin{tcolorbox}[boxsep=1pt,left=2pt,right=2pt,top=2pt,bottom=2pt, boxrule=0mm]
\textbf{Highlight-3:} The efficiency benefit of \texttt{PriBOM} is widely endorsed for its ability to streamline workflows, facilitate swift onboarding, bridge communication gaps between diverse roles, and enhance both efficiency and privacy environment. However, the feedback highlights a need for ongoing refinements to ensure its practical adaptation and alignment with real-world demands across different roles.
\end{tcolorbox}

\textbf{Future Enhancements.} Participants also proposed additional information to include in \texttt{PriBOM} in different use cases. For example, one product manager suggested the potential inclusion of ``\textit{security level for guarding PII fields}''. 
This information may benefit the implementation of more targeted, effective data protection strategies based on data sensitivity and decision-making.
Another valuable suggestion from a project manager is ``\textit{Data about TPL should include github or web page of the project}'', facilitating quick investigation and smooth work in due diligence and verifying the credibility and security practices of TPLs.

Current findings reveal the promising potential of \texttt{PriBOM} to effectively bridge the gap between various roles, enhancing communication, compliance, and efficiency within development teams. Nonetheless, the deployment of the \texttt{PriBOM} approach within diverse environments underscores the necessity for further efforts to refine its adaptability and to better meet the specific real-world needs highlighted by various roles. 

\section{Limitations}
\label{sec_limitation}

While we propose a pre-fill of \texttt{PriBOM} and access the usefulness, there are still limitations. First, the pre-fill generation is highly dependent on the performance of static analysis tools, and limited performance could result in restricted quality of pre-fill. Second, our user study design can be improved by involving a in-depth interview to mitigate the possibility of social desirability or acquiescence biases. Although we propose this promising privacy enhancing mechanism in software development process, real practicability needs to be further verified. Still, as the first of such kind, we believe this work can ignite deeper thinking on the aspect of software engineering at the systematic level, considering all practitioners to solve the long-standing privacy policy challenges.


\section{Conclusion}
\label{sec_conclusion}

Privacy regulations commonly require developers to provide authentic and comprehensive privacy notices, e.g., privacy policies or labels, in communicating their apps’ privacy practices. 
However, developers often struggle to create and maintain accurate privacy notices, especially for sophisticated apps with complex features and large development teams.
We introduce \texttt{PriBOM} (Privacy Bills of Materials) as a systematic and collaborative solution for the development team to craft accurate privacy notices.
As a privacy information inventory, we leverage static analysis and privacy notice analysis techniques to implement \texttt{PriBOM}.
The role of \texttt{PriBOM} in enhancing privacy-related communication is well received with 83.33\% agreement, underscoring the usefulness and practicability of \texttt{PriBOM} in fostering a privacy-conscious development workflow.
Lastly, we discuss in depth the implications of our results, including differences in views held by different roles regarding \texttt{PriBOM} and privacy notice generation, especially non-technical roles.

\bibliographystyle{ACM-Reference-Format}
\bibliography{8_References}

\section{Appendix}

\subsection{The Methodology and Scope of Literature in Formative Study}

Table~\ref{tab:developer_study} presents the methodology and scope of studies on privacy-related challenges faced by developers. The most common methodology is the semi-structured interview, which is used in over half of the listed studies. Other studies utilize qualitative methods such as email interactions and analysis of online community discussions. A rich tapestry of collected data reflects the multifaceted nature of privacy challenges in the development process. Participant backgrounds vary widely, encompassing a spectrum from Android and iOS app developers to users of online communities like Stack Overflow and Reddit. This variety underscores the breadth of privacy concerns across different platforms and development environments. It is worth noting that participants of some studies include project managers and legal teams, emphasizing the cross role nature of addressing privacy issues in software development and reflecting the interaction between technical and non-technical factors in privacy management.

%
%
\begin{table*}[!t]
\centering
  \caption{Methodology and scope of studies on privacy-related challenges faced by developers.}
  \label{tab:developer_study}
  \resizebox{0.9\linewidth}{!}{%
  \begin{tabular}{l | l | l}
  \toprule
    Study & Methodologies & Participant Background  \\
    \midrule
    Khandelwa et al.~\cite{khandelwal2023unpacking} & Qualitative study involving email interactions & More than 3,500 Android app developers\\ \midrule
    Li et al.~\cite{li2022understanding} & \makecell[l]{Remotely observational study on Zoom,\\ semi-structured interview} & 12 iOS app developers from various platforms\\ \midrule
    Li et al.~\cite{li2018coconut} & Semi-structured interview & \makecell[l]{9 Android app developers, including independent and full-time \\developers, researchers, and others}\\ \midrule
    Li et al.~\cite{li2021developers} & \makecell[l]{Qualitative analysis of discussions \\from the /r/androiddev subreddit} & Users of the /r/androiddev subreddit\\ \midrule
    Balebako et al.~\cite{balebako2014privacy} & Semi-structured interview, online survey & \makecell[l]{13 app developers varied in terms of company size and app types,\\ 228 U.S. app developers and product managers}\\ \midrule
    Lee et al.~\cite{lee2024don} & Semi-structured interview & \makecell[l]{35 industry practitioners, including researchers,\\ software engineers, and designers}\\ \midrule
    Seymour et al.~\cite{seymour2023voice} & Semi-structured interview & \makecell[l]{30 developers varied in terms of experience and professionals}\\ \midrule
    Weir et al.~\cite{weir2020needs} & Online Survey & \makecell[l]{345 Google Play Android developers}\\ \midrule
    Kek{\"u}ll{\"u}o{\u{g}}lu et al.~\cite{kekulluouglu2023we} & Semi-structured interview & \makecell[l]{16 developers in Turkish software startups}\\ \midrule
    Tahaei et al.~\cite{tahaei2021privacy} & Semi-structured interview & \makecell[l]{12 Privacy Champions who promote privacy in development teams}\\ \midrule
    Tahaei et al.~\cite{tahaei2020understanding} & \makecell[l]{Qualitative analysis of privacy-related \\questions from Stack Overflow} & \makecell[l]{Users of Stack Overflow}\\ \bottomrule

\end{tabular}
}%
\end{table*}

\subsection{Challenges for Legal Experts}

%
\begin{figure} [h]
\begin{subfigure}{.9\linewidth}
  \centering
  \includegraphics[width=.99\linewidth]{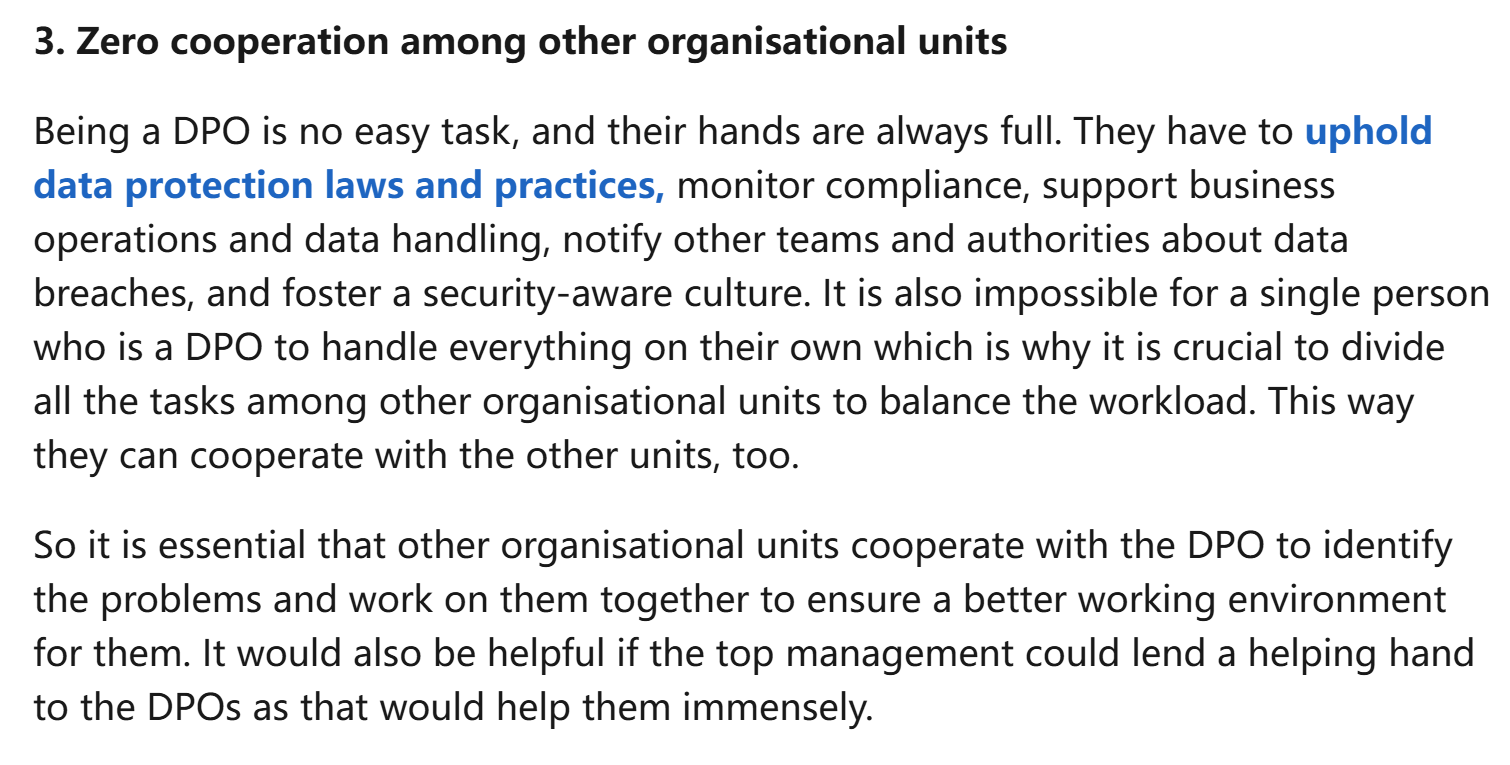}
  \label{fig_DPO_2}
\end{subfigure}%
\hfill
\begin{subfigure}{.9\linewidth}
  \centering
  \includegraphics[width=.99\linewidth]{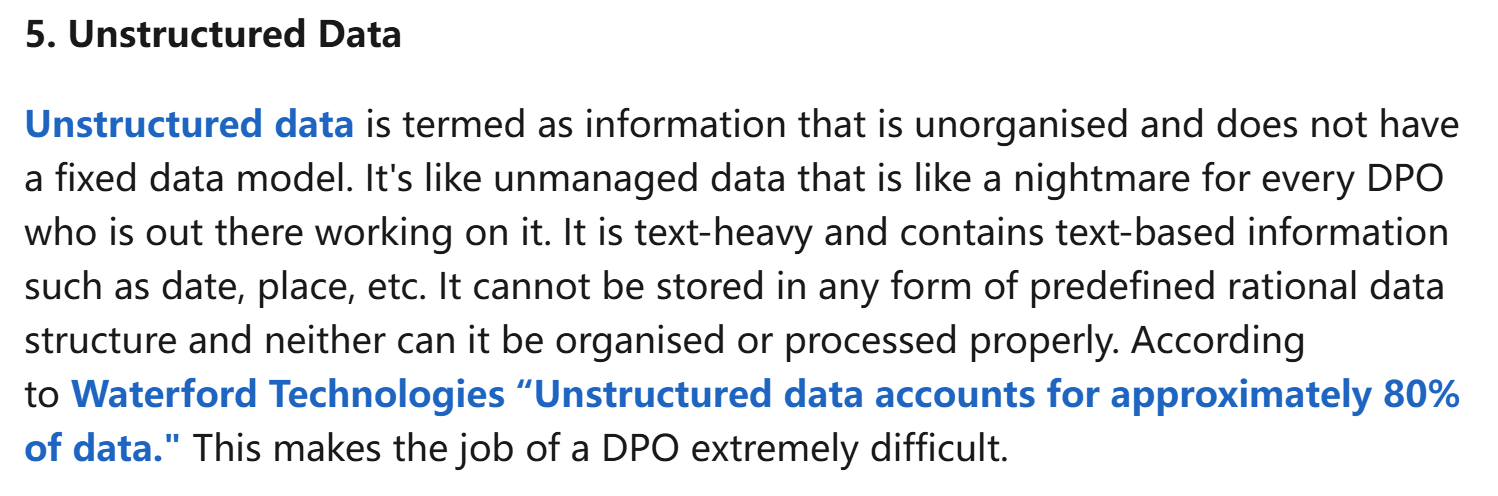}
  \label{fig_DPO_3}
\end{subfigure}
\hfill
\begin{subfigure}{.9\linewidth}
  \centering
  \includegraphics[width=.99\linewidth]{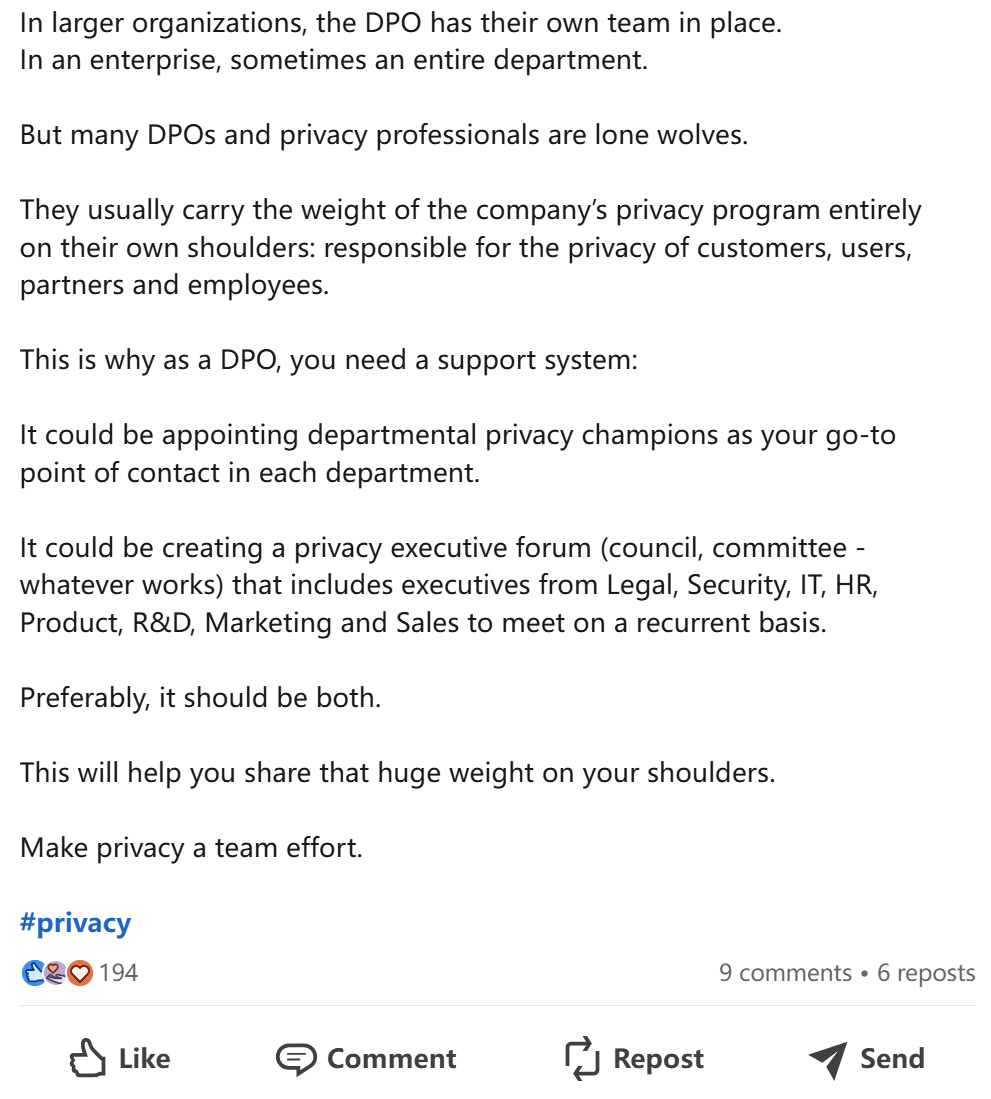}
  \label{fig_DPO_1}
\end{subfigure}
\begin{subfigure}{.9\linewidth}
  \centering
  \includegraphics[width=.99\linewidth]{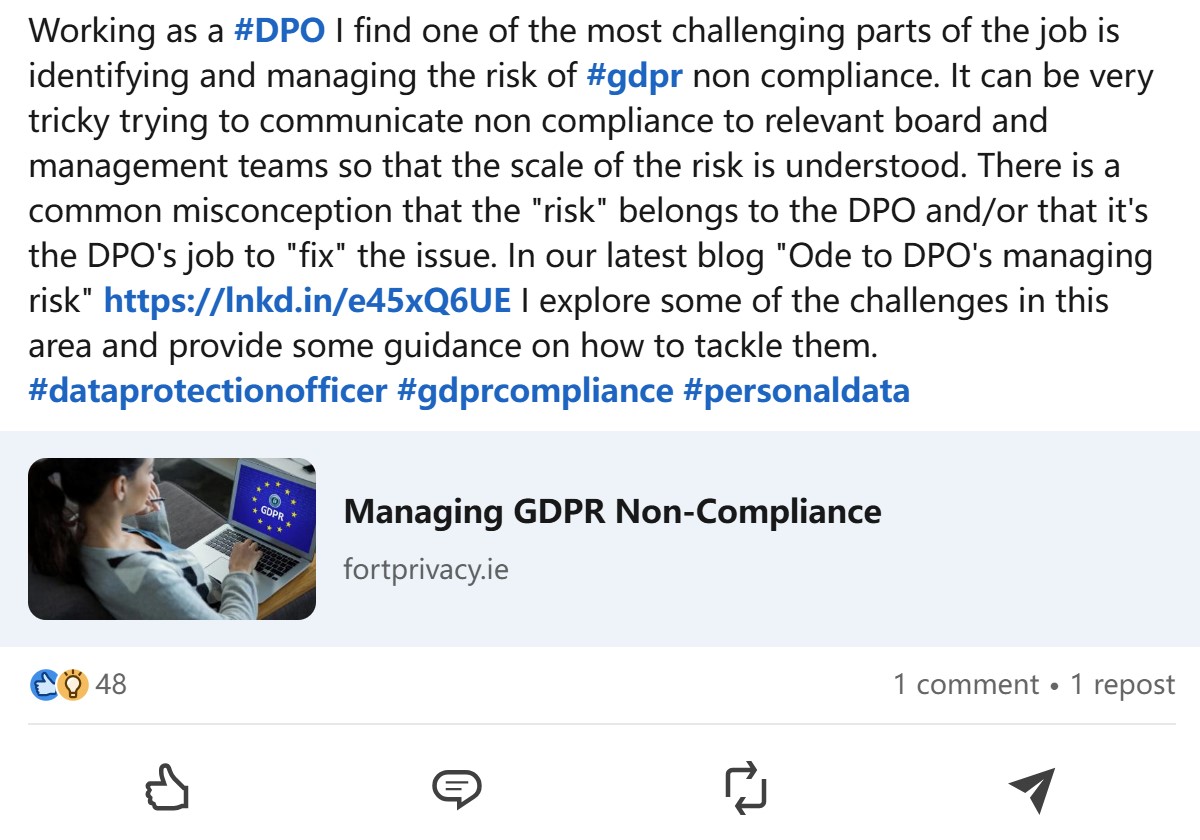}
  \label{fig_DPO_4}
\end{subfigure}
\caption[Caption]{The screenshots of data protection officers' posts on LinkedIn~\cite{LinkedIn, LinkedIn2, LinkedIn3}.}
\label{fig_DPO}
\end{figure}
%

As shown in Figure~\ref{fig_DPO}, legal experts like DPO in enterprises emphasized the severe challenges they face, including various factors such as the lack of resources, zero cooperation among organisational units, and unstructured data. Among them, the necessity for support from other organizational units is highlighted, underscoring the importance of distributing privacy tasks across various development roles to balance the workload effectively. \texttt{PriBOM} serves as a systematic approach to liberating legal experts from heavy and difficult tasks.

\end{document}